\documentclass[twoside,english]{aastex6}
\usepackage[T1]{fontenc}
\usepackage[latin9]{inputenc}
\usepackage{array}
\usepackage{float}
\usepackage{multirow}
\usepackage{amsbsy}
\usepackage{amstext}
\usepackage{amssymb}
\usepackage{graphicx}
\usepackage{setspace}

\makeatletter

\providecommand{\tabularnewline}{\\}

\usepackage{amsmath}
\slugcomment{}
\shorttitle{}
\shortauthors{}

\makeatother

\usepackage{babel}
\begin{document}

\title{Cosmological studies from tomographic weak lensing peak abundances
and impacts of photo-z errors}

\author{Shuo Yuan\altaffilmark{1}, Chuzhong Pan\altaffilmark{1}, Xiangkun
Liu\altaffilmark{2},Qiao Wang\altaffilmark{3}, Zuhui Fan\altaffilmark{2,1}}

\altaffiltext{1}{Department of Astronomy, School of Physics, Peking University, Beijing
100871, China;yuanshuoastro@gmail.com; fanzuhui@pku.edu.cn}

\altaffiltext{2}{South-Western Institute for Astronomy Research, Yunnan University
Chenggong District, Kunming 650500, Yunnan Province, China; zuhuifan@ynu.edu.cn}

\altaffiltext{3}{Key Laboratory of Computational Astrophysics, National Astronomical
Observatories, Chinese Academy of Sciences, Beijing, 100012, China;}
\begin{abstract}

Weak lensing peak abundance analyses have been applied in different
surveys and demonstrated to be a powerful statistics in extracting
cosmological information complementary to cosmic shear two-point correlation
studies. Future large surveys with high number densities of galaxies
enable tomographic peak analyses. Focusing on high peaks, we investigate
quantitatively how the tomographic redshift binning can enhance the
cosmological gains. We also perform detailed studies about the degradation
of cosmological information due to photometric redshift (photo-z)
errors. We show that for surveys with the number density of galaxies
$\sim40\,{\rm arcmin^{-2}}$, the median redshift $\sim1$, and the
survey area of $\sim15000\,{\rm deg^{2}}$, the 4-bin tomographic
peak analyses can reduce the error contours of $(\Omega_{{\rm m}},\sigma_{8})$
by a factor of $5$ comparing to 2-D peak analyses in the ideal case
of photo-z error being absent. More redshift bins can hardly lead
to significantly better constraints. The photo-z error model here
is parametrized by $z_{{\rm bias}}$ and $\sigma_{{\rm ph}}$ and
the fiducial values of $z_{{\rm bias}}=0.003$ and $\sigma_{{\rm ph}}=0.02$
is taken. We find that using tomographic peak analyses can constrain
the photo-z errors simultaneously with cosmological parameters. For
4-bin analyses, we can obtain $\sigma(z_{{\rm bias}})/z_{{\rm bias}}\sim10\%$
and $\sigma(\sigma_{{\rm ph}})/\sigma_{{\rm ph}}\sim5\%$ without
assuming priors on them. Accordingly, the cosmological constraints
on $\Omega_{{\rm m}}$ and $\sigma_{8}$ degrade by a factor of $\sim2.2$
and $\sim1.8$, respectively, with respect to zero uncertainties on
photo-z parameters. We find that the uncertainty of $z_{{\rm bias}}$
plays more significant roles in degrading the cosmological constraints
than that of $\sigma_{{\rm ph}}$.

\end{abstract}


\section{Introduction}

Weak gravitational lensing (WL) effects have been shown to be a powerful
probe in cosmological studies \citep{2008A&A...479....9F,2011MNRAS.411.2161M,2013MNRAS.432.2433H,2013MNRAS.433.2545E,2014MNRAS.441.2725F,2017MNRAS.465.1454H,2016PhRvD..94b2001A,2019PASJ..tmp...22H}.
Their high signals come dominantly from nonlinear regions of large-scale
structures. Therefore to fully explore the cosmological information
embedded in the WL data, different statistical analyses are needed.
Besides cosmic shear two point correlation (2PCF) studies, WL peak
abundance analyses have emerged to be an important statistics to extract
cosmological information complementary to 2PCF \citep{2014MNRAS.442.2534S,2015MNRAS.450.2888L,2018MNRAS.474..712M,2018MNRAS.474.1116S,2015PhRvD..91f3507L,2010MNRAS.402.1049D,2014IAUS..306..107L}.

Theoretically, WL effects depend on the formation and evolution of
large-scale structures, and on the cosmic expansion history through
the lensing efficiency kernel \citep{2001PhR...340..291B}. Thus source
galaxies at different redshifts experience different lensing effects,
and tomographic analyses by dividing source galaxies into different
redshift bins can significantly enhance the cosmological gains comparing
to the 2-D studies without binning \citep{1999ApJ...522L..21H}. Tomographic
2PCF analyses have been extensively studied and applied to different
WL surveys \citep{2008A&A...479....9F,2013MNRAS.433.2545E,2017MNRAS.465.1454H,2016PhRvD..94b2001A,2019PASJ..tmp...22H}.
For WL peak statistics, limited to the relatively low number density
of source galaxies from current surveys, tomographic studies have
not been applied to WL data analyses. However, future large surveys
will be able to produce much larger samples with the galaxy number
reaching $\sim10^{9}$ \citep{2009arXiv0912.0201L,2011SSPMA..41.1441Z,2018LRR....21....2A,2018arXiv180810458H},
and thus enable the tomographic WL peak studies. From simulations,
a number of studies have been carried out to explore the benefit of
employing tomographic WL peak analyses (\citealt{2010MNRAS.402.1049D,2014IAUS..306..107L,2015PhRvD..91f3507L,2016PhRvD..94f3534P,2018MNRAS.474..712M,2018arXiv181001781L,2018arXiv181204018P}).

For WL surveys, practically, the galaxy redshifts can only be derived
from multi-band photometric observations. Such photometric redshifts
(photo-z) inevitably have errors. Therefore for tomographic WL analyses,
to assess the impacts of photo-z errors on the derived cosmological
constraints is crucially important. From these studies, we can also
set requirements for the accuracy of photo-z estimates and the subsequent
calibrations \citep{2006ApJ...636...21M,2006MNRAS.366..101H,2008ApJ...682...39M,2008MNRAS.391..228A,2009ApJ...699..958S,2010MNRAS.401.1399B,2010ApJ...720.1351H,2012JCAP...04..034H,2017JCAP...10..056Y,2018MNRAS.480.2178C}.
While the impacts of the photo-z errors on the tomographic 2PCF have
been systematically studied in depth \citep{2008ApJ...682...39M,2008MNRAS.386..781M,2018arXiv180810458H},
similar analyses for WL peak statistics are still lacking. In \citet{2018arXiv181012312A},
they investigated the influences of photo-z errors on tomographic
peak analyses from simulations. In their studies, the cosmology dependence
of peak abundances is solely from simulations spanning a range of
cosmological parameters. Thus for each value of photo-z errors, a
separate WL simulation is needed for each of the cosmological models
considered. It is therefore not easy for them to study the capability
of using tomographic analyses to constrain the photo-z error parameters
simultaneously with cosmological parameters, and the corresponding
degradations on cosmological constraints and the requirements for
the prior knowledge on photo-z errors. Instead, they studied the cosmological
parameter bias induced from different photo-z biases, and used the
B/U ratio (bias vs. statistical uncertainty) for a cosmological parameter
as an estimate for its degradation factor.

In this paper, we carry out the studies on tomographic peak analyses.
We focus on high WL peaks, and employ our theoretical model of \citet{2018ApJ...857..112Y}
for calculating the cosmological dependence of tomographic peak abundances.
This model is an extension of the model of \citet{2010ApJ...719.1408F}
and includes both the shape noise and the projection of large-scale
structures into consideration. With this theoretical basis, we are
able to investigate the cosmological gains of different redshift binning
of tomographic peak analyses and the impact of the photo-z errors.
For the latter, we investigate the capability of simultaneously constraining
the photo-z error parameters and the cosmological parameters using
tomographic peak abundance data, and the degradation of the cosmological
constraints due to the propagation of photo-z errors into the uncertainties
of cosmological parameters . We then further examine the requirements
of the prior knowledge on the photo-z error parameters with respect
to the degradation requirements. To facilitate and validate our analyses,
we also use large ray tracing simulations to generate mock data taking
into account the photo-z errors in our studies.

The rest of the paper is organized as follows. In Sec.2, we summarize
our theoretical model of \citet{2018ApJ...857..112Y} for high WL
peak abundances. In Sec.3, we describe our simulations. In Sec. 4,
we show the cosmological gains of tomographic peak analyses with different
redshift binnings in the ideal case without photo-z errors. We present
our detailed analyses on the impact of photo-z errors in Sec.5. Conclusions
and discussions are shown in Sec. 6.

\section{THE MODEL FOR WEAK LENSING HIGH PEAK ABUNDANCES}

The WL effect arises from the light deflection by large-scale structures
in the Universe. The induced observational effects can be described
by the second derivatives of the lensing potential $\phi$, i.e.,
the Jacobin matrix given by \citep{2001PhR...340..291B}:

\begin{equation}
\ensuremath{\mathbf{A}=\left(\begin{array}{cc}
{1-\kappa-\gamma_{1}} & {-\gamma_{2}}\\
{-\gamma_{2}} & {1-\kappa+\gamma_{1}}
\end{array}\right)}
\end{equation}
where

\begin{equation}
\begin{array}{c}
{\kappa=\frac{1}{2}\nabla^{2}\phi;\quad\gamma_{1}=\frac{1}{2}\left(\frac{\partial^{2}\phi}{\partial^{2}x_{1}}-\frac{\partial^{2}\phi}{\partial^{2}x_{2}}\right),\quad\gamma_{2}=\frac{\partial^{2}\phi}{\partial x_{1}\partial x_{2}},}\end{array}
\end{equation}
with $\mathbf{x}=(x_{1},x_{2})$ being the two-dimensional angular
vector.

The convergence $\kappa$, reflecting the isotropic change of a background
image, is related to the projected density fluctuation weighted by
the lensing kernel under the Born approximation, and thus can intuitively
reveal the (dark) matter distribution in the Universe. The $\gamma$
components lead to anisotropic shears to an image. Observationally,
WL signals are extracted by measuring accurately the shapes of background
galaxies, and thus statistically directly related to the reduced shear
defined as $\ensuremath{g_{i}=\gamma_{i}/(1-\kappa)}$. Because of
the physical relation between $\kappa$ and $\gamma$, the convergence
field $\kappa$ can be reconstructed from the galaxy shape measurements,
which inevitably includes the contamination from the shape noise resulting
from the intrinsic ellipticities of source galaxies. We can also construct
a scalar aperture mass field $M_{{\rm ap}}$ directly from the reduced
shears, and thus avoiding the possible artificial effects from the
$\kappa$ reconstructions. In WL lensing regime with $\kappa\ll1$,
and $\ensuremath{g_{i}\approx\gamma_{i}}$ , the $M_{{\rm ap}}$ field
is approximately the same as the convergence field convolved with
a compensated filter.

In this study, we aim to analyze the benefit from tomographic WL peak
analyses, and how the photo-z errors affect the cosmological studies.
For that, we concentrate on high peaks in the convergence field smoothed
with a Gaussian kernel. We note, however, that the peak model to be
described is also applicable to the convergence field filtered by
compensated kernels \citep{2018ApJ...857..112Y}. Moreover, following
the same halo approach, we have developed a theoretical model for
high peaks in $M_{{\rm ap}}$ constructed directly from the reduced
shear field, and the results agree with that from simulations very
well (Pan et al. in preparation). Therefore the methodologies shown
in this paper can be readily extended to investigate tomographic $M_{{\rm ap}}$
peaks.

For a high convergence peak, studies have shown that its signal is
typically dominated by the contribution from a single massive halo
along the line of sight \citep{2004MNRAS.350..893H,2011PhRvD..84d3529Y,2016PhRvD..94d3533L,2018MNRAS.478.2987W}.
This leads to the halo-based model of \citet{2010ApJ...719.1408F}
for WL high peaks, in which the Gaussian shape noise is taken into
account. In \citet{2018ApJ...857..112Y}, we extended the model by
further including the projection effect from large-scale structures
in the calculation. This improvement is important because future surveys
will go deeper and thus the projection effect can be comparable to
that from the shape noise. In this model, we consider halo regions
and the field region, separately. In halo regions, the smoothed convergence
field can be written as

\begin{equation}
\mathcal{K}=\mathcal{K}_{\mathrm{H}}+\mathcal{K}_{\mathrm{LSS}}+\mathcal{N},\label{eq:start_point}
\end{equation}
where $\mathcal{K}_{\mathrm{H}}$ is the contribution from massive
halos with mass larger than $M_{*}$, $\mathcal{K}_{\mathrm{LSS}}$
is the contribution from the projection effect of large-scale structures
excluding those from the massive halos already considered, and $\mathcal{N}$
is from the shape noise. Both $\mathcal{K}_{\mathrm{LSS}}$ and $\mathcal{N}$
are modeled as Gaussian random fields. For the shape noise field $\mathcal{N}$,
the moments are given by

\begin{equation}
\sigma_{\mathrm{N},i}^{2}=\int_{0}^{\infty}\frac{\ell\mathrm{d}\ell}{2\pi}\ell^{2i}C_{\ell}^{\mathrm{N}}\,(i=0,1,2),\label{eq:3moments}
\end{equation}
with $C_{\ell}^{\mathrm{N}}$ being the power spectrum of the smoothed
noise field. For $\mathcal{K}_{\mathrm{LSS}}$ , we have

\begin{equation}
\sigma_{\mathrm{LSS},i}^{2}=\int_{0}^{\infty}\frac{\ell\mathrm{d}\ell}{2\pi}\ell^{2i}C_{\ell}^{\mathrm{LSS}};(i=0,1,2)
\end{equation}
where the power spectrum $C_{\ell}^{\mathrm{LSS}}$ is calculated
by

\begin{equation}
\ensuremath{C_{\ell}=\frac{9H_{0}^{4}\Omega_{m}^{2}}{4}\int_{0}^{\chi_{H}}\mathrm{d}\chi^{\prime}\frac{w^{2}\left(\chi^{\prime}\right)}{a^{2}\left(\chi^{\prime}\right)}P_{\delta}^{\mathrm{LSS}}\left(\frac{\ell}{f_{K}\left(\chi^{\prime}\right)},\chi^{\prime}\right)}\label{eq:WLsignal}
\end{equation}
where $a$ is the cosmic scale factor, and $\ensuremath{w\left(\chi^{\prime}\right)}$
is the lensing kernel function

\begin{equation}
\ensuremath{w\left(\chi^{\prime}\right)=\int_{\chi^{\prime}}^{\chi_{H}}\mathrm{d}\chi p_{\mathrm{s}}(\chi)\frac{f_{K}\left(\chi-\chi^{\prime}\right)}{f_{K}(\chi)}}.\label{eq:WLeff}
\end{equation}

Here $p_{\mathrm{s}}$ is the source redshift distribution function.
For $P_{\delta}^{\mathrm{LSS}}$, in \citealp{2018ApJ...857..112Y},
it is modeled by subtracting the one halo term from massive halos
with mass above the threshold $M_{*}$ from the full non-linear power
spectrum, i.e., (For more details please see Eq.(18)-Eq.(23) in \citealt{2018ApJ...857..112Y}.)

\begin{equation}
P_{\delta}^{\mathrm{LSS}}[k,\chi(z)]=P_{\delta}[k,\chi(z)]-\left.P_{\delta}^{1\mathrm{H}}\right|_{M\geqslant M_{*}}[k,\chi(z)].
\end{equation}
Finally for the combined Gaussian random field, $\mathcal{K}_{\mathrm{LSS}}+\mathcal{N},$
its moments are given by

\begin{equation}
\sigma_{i}^{2}=\sigma_{\mathrm{LSS},i}^{2}+\sigma_{\mathrm{N},i}^{2}.\ (i=0,1,2)
\end{equation}

Then for an individual halo region, we can calculate the peak distribution
using the Gaussian random field theory modulated by the halo profile
\citep{2010ApJ...719.1408F}. It is noted that in this region, it
contains the peak corresponding to the original halo peak but the
position and the height are changed due to the existence of the combined
Gaussian random field. It also has peaks from the Gaussian random
field with the heights modulated by the halo profile. For peaks in
all the massive halo regions, we integrate over the massive halos
with the number weighted by the halo mass function.

The field region corresponds to regions outside halos considered above.
In this region, the peak distribution can be calculated from the combined
Gaussian random field $\mathcal{K}_{\mathrm{LSS}}+\mathcal{N}$ without
halo modulations. The total peak abundance is then computed by the
summation of peaks in halo regions and that in the field region \citep{2018ApJ...857..112Y}.

In the next section, we describe our tomographic simulations, and
also compare the simulation results with the model predictions to
validate the applicability of the model to tomographic WL peak analyses.

\section{Simulated convergence maps}

The same 24 sets of $N$-body simulation data as in \citealp{2015MNRAS.450.2888L}
and in \citealp{2018ApJ...857..112Y} are used here. They are under
the flat $\Lambda$CDM model with the cosmological parameters $(\Omega_{m},\Omega_{\Lambda},\Omega_{b},h,n_{s},\sigma_{8})$
= (0.28, 0.72, 0.046, 0.7, 0.96, 0.82). Each set consists of 12 independent
simulation boxes, with 8 having the box size of 320 $h^{-1}{\rm Mpc}$
and the particle number of $600^{3}$ to fill the region from $z=0$
to $z=1$, and 4 larger ones with the same particle number but the
size of 600 $h^{-1}{\rm Mpc}$ for the region with $1<z\leq3$. From
each set of simulations, the WL ray tracing calculations are done
using 59 lens planes up to $z=3$. For each plane, we store the convergence
and shear data computed from the lens planes before it. The convergence
maps with different redshift distributions can then be constructed
from these 59 maps by weighting each plane according to the considered
source redshift distribution function $p(z)$. Specifically to generate
2-D convergence maps without tomography, we use,

\begin{equation}
\ensuremath{\kappa_{\mathrm{mock}}(\boldsymbol{\theta})=\frac{1}{A}\sum_{{\rm s}=1}^{59}p\left(z_{\mathrm{s}}\right)\kappa\left(\boldsymbol{\theta};z_{\mathrm{s}}\right)\left(z_{\mathrm{s}+1}-z_{\mathrm{s}}\right)}\label{eq:kappa_mock}
\end{equation}
where $A$ is the normalization factor for $p(z)$. In this paper,
we adopt the overall redshift distribution as follows \citep{2009arXiv0912.0201L}

\begin{equation}
\ensuremath{p(z)=n_{{\rm g0}}\left(\frac{z}{z_{0}}\right)^{2}\exp\left(-\frac{z}{z_{0}}\right)}\label{eq:prob_z_LSST}
\end{equation}
from $z=0$ to $z=3.0$ and $z_{0}=0.3$ mimicking the LSST-like surveys.
From each of the 24 sets of simulations, we can obtain 4 maps each
with the area of $3.5\times3.5$ deg$^{2}$ pixelized into $1024\times1024$
data points. Thus totally we have $24\times4\times3.52=1176$ deg$^{2}$
convergence data for each redshift distribution considered. We then
add the Gaussian shape noise to each pixel according to

\begin{equation}
\sigma_{\mathrm{pix}}^{2}=\frac{\sigma_{\epsilon}^{2}}{2n_{{\rm g}}\theta_{\mathrm{pix}}^{2}},\label{eq:pixnoise}
\end{equation}
where we take $\sigma_{\epsilon}=0.4$. \textbf{It is noted that this
$\sigma_{\epsilon}$ is the dispersion of the total ellipticities
including two components. Its value is related to both the intrinsic
ellipticity distribution of source galaxies, and the galaxy shape
measurement errors. The value of $\sim0.4$ taken here is in accord
with that of CFHTLenS observations \citep{2013MNRAS.430.2200K}. }And
the pixel size of maps $\ensuremath{\theta_{\mathrm{pix}}=3.5\times60/1024=0.205}$
arcmin. In the 2-D case, the source galaxy number density $n_{{\rm g0}}$
is taken to be 40 arcmin$^{-2}$. We then apply a Gaussian smoothing
with the kernel given by

\begin{equation}
W_{\theta_{G}}(\boldsymbol{\theta})=\frac{1}{\pi\theta_{G}^{2}}\exp\left(-\frac{|\boldsymbol{\theta}|^{2}}{\theta_{G}^{2}}\right).
\end{equation}
We take $\theta_{G}$ = 2.0 arcmin to obtain the final smoothed noisy
convergence maps.

From the smoothed convergence maps, we identify peaks if their convergence
values are larger than those of the surrounding 8 pixels. Then we
record the peak height scaled by the shape noise $\sigma_{N,0}$.
To avoid the possible boundary effects on the smoothed maps, we exclude
the outer 70 pixels along each side of a map, which corresponds to
about $7\times\theta_{G}$ in the peak analyses. Thus the total effective
area used in our analyses is about 876 deg$^{2}$.

To simulate the tomographic convergence maps, in the ideal case without
photo-z errors, we divide the source galaxies into different bins.
For the bin with $z\in[a,b]$, the $p(z)$ and $n_{g}$ used in Eq.\eqref{eq:kappa_mock}
and Eq.\eqref{eq:pixnoise} are computed by

\begin{equation}
p(z)=\begin{cases}
\left(\frac{z}{z_{0}}\right)^{2}\exp\left(-\frac{z}{z_{0}}\right); & z\in[a,b]\\
0. & z\notin[a,b]
\end{cases}
\end{equation}
and
\begin{equation}
n_{g}(z\in[a,b])=\frac{\int_{a}^{b}{\rm d}z\,p(z)}{\int_{0}^{3}{\rm d}z\,p(z)}\cdot n_{g0}.\label{eq:ng_tombin}
\end{equation}
The corresponding normalization factor is $\ensuremath{A_{\mathrm{ab}}=\int_{a}^{b}\mathrm{d}z\,p(z)}$.

\begin{figure}[h]
\begin{centering}
\includegraphics[scale=0.35]{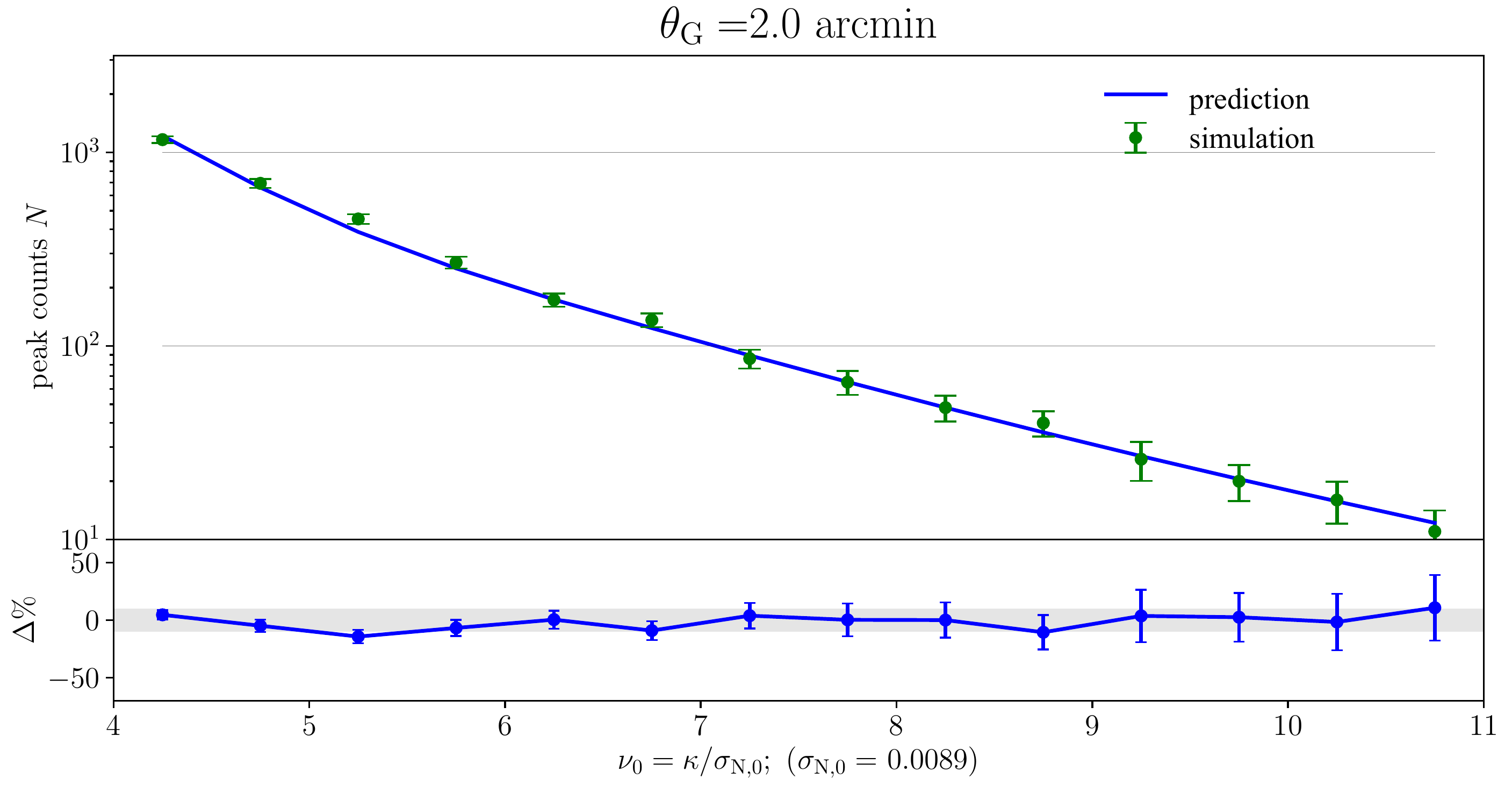}
\par\end{centering}
\figcaption{\label{fig:wihtout_tom}Comparison of peak abundance from simulations
(green points with error bars) with that from our model prediction
(blue line) in the 2-D case without tomography. The redshift distribution
is from Eq.\eqref{eq:prob_z_LSST}, and the number density of galaxies
is $n_{g}=40\ \hbox{arcmin}^{-2}$. The lower panel shows the relative
differences between the theoretical results and the simulation results
\textbf{and the gray regions are $\pm10\%$ range.}}
\end{figure}

\begin{figure}[H]
\begin{centering}
\includegraphics[scale=0.3]{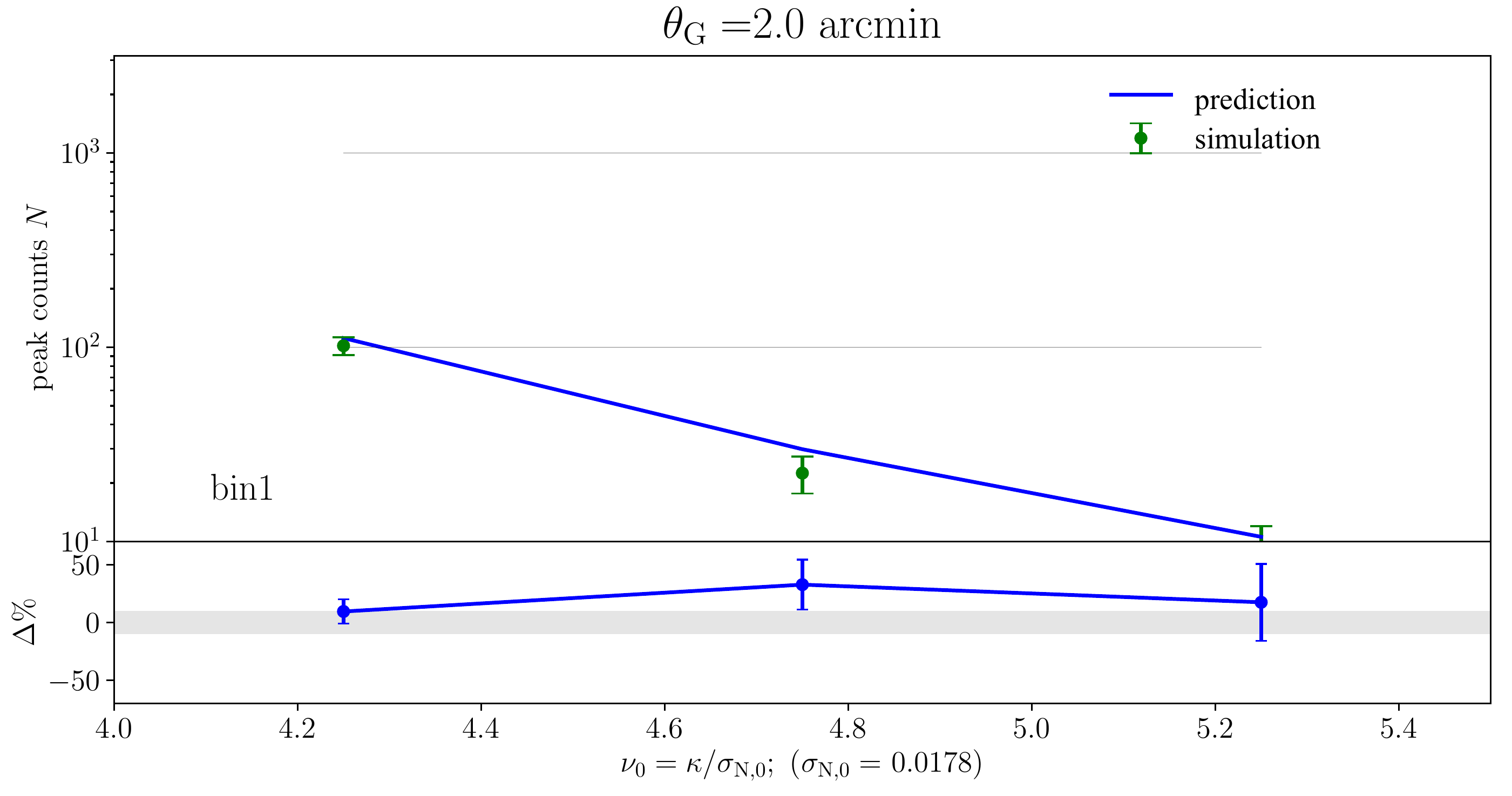}\includegraphics[scale=0.3]{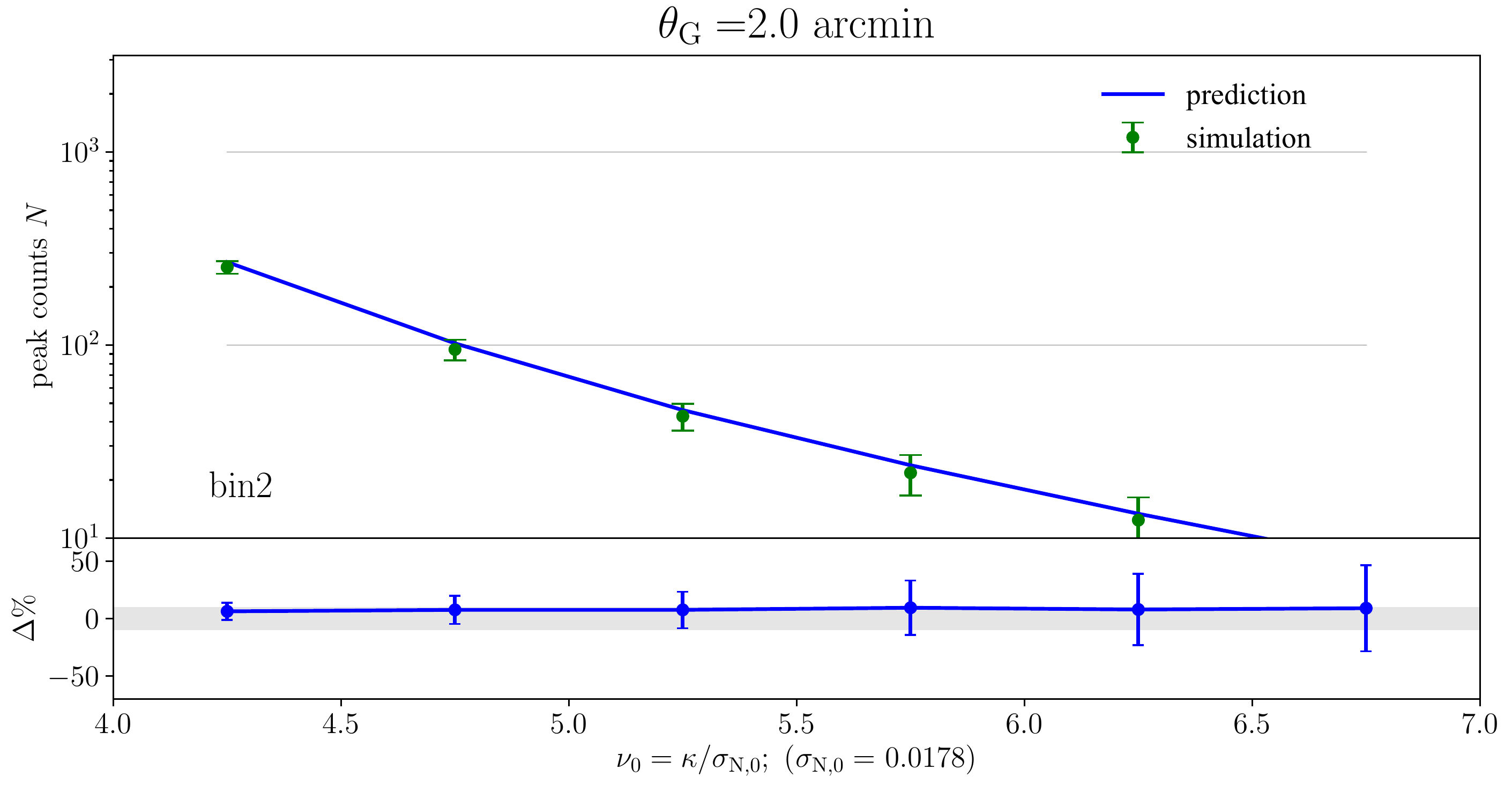}
\par\end{centering}
\begin{centering}
\includegraphics[scale=0.3]{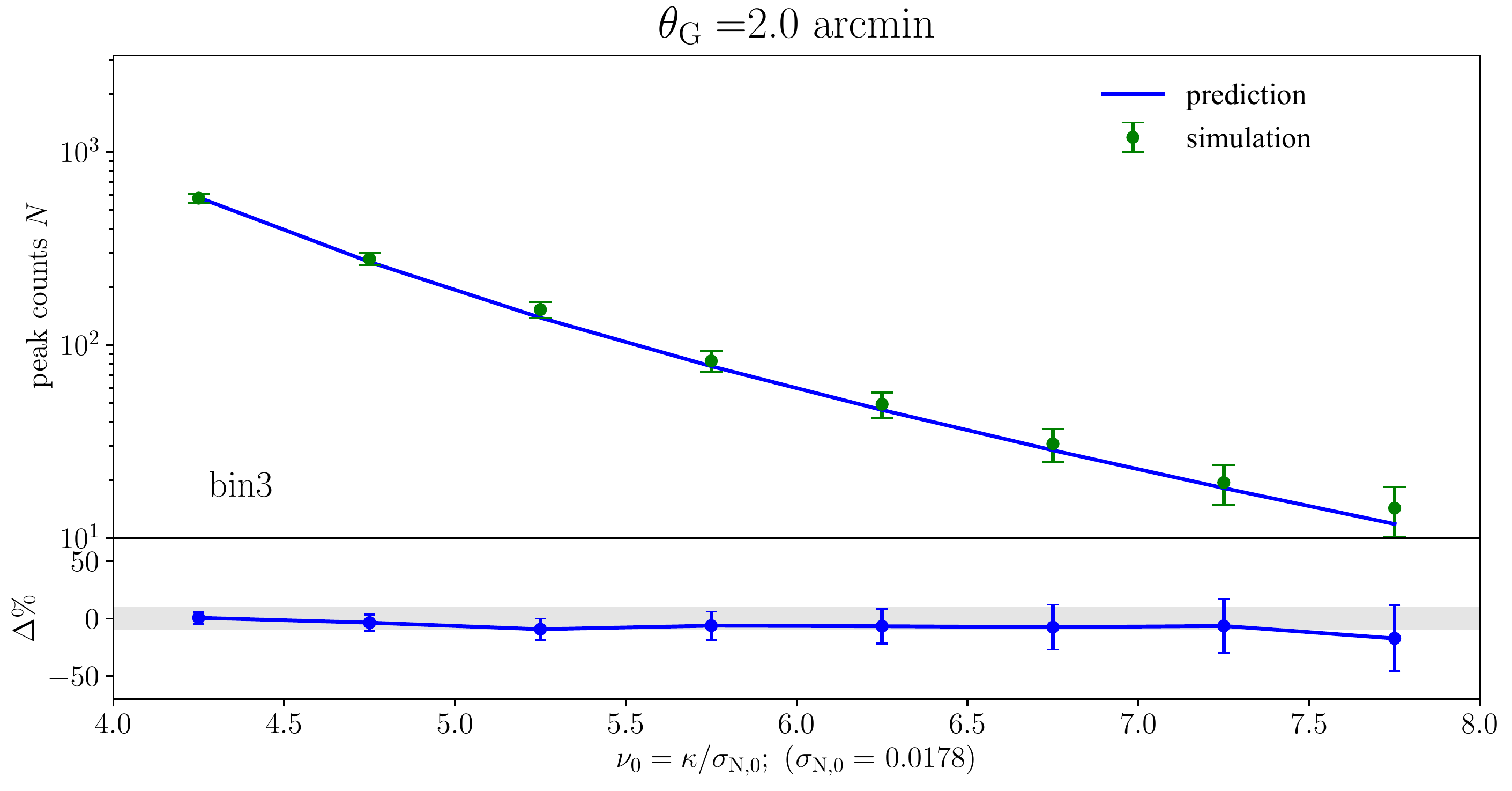}\includegraphics[scale=0.3]{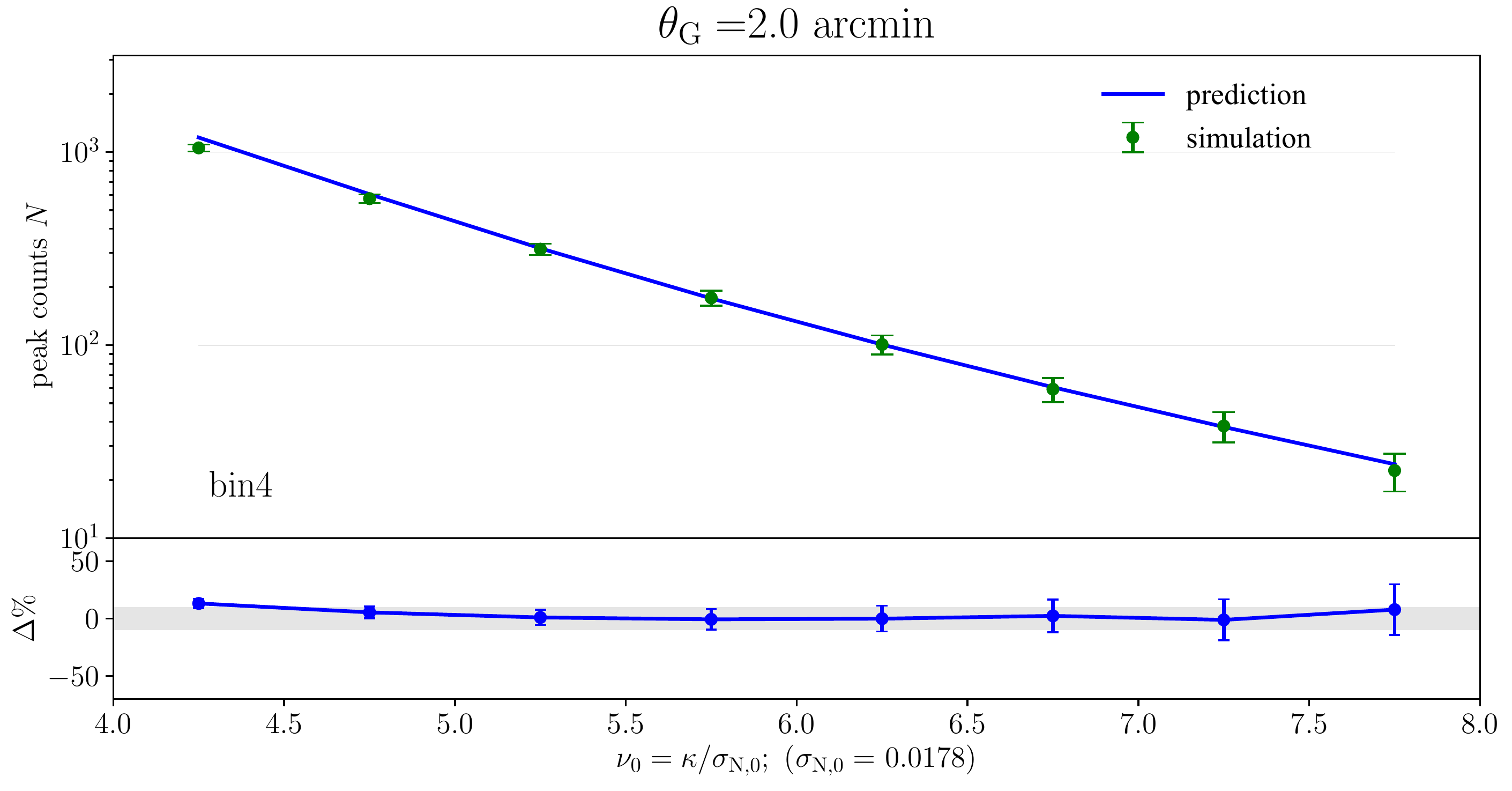}
\par\end{centering}
\figcaption{\label{fig:4bins}Similar to Fig.\ref{fig:wihtout_tom}, but for the
4-bin tomographic peak abundance. The number density of source galaxies
is $n_{g}=10\hbox{arcmin}^{-2}$ for each bin.}
\end{figure}

In Fig.\ref{fig:wihtout_tom}, we show the peak abundances in the
2-D case. \textbf{The blue line is our theoretical prediction, and
the green points with error bars are from simulations. For illustration
purposes, the error bars here show the Poisson errors for the number
of peaks in different bins in the total simulated area of $\sim876\deg^{2}$,
without considering the covariance between different bins. In the
later cosmological analyses in this work, we take into account the
full covariance. }The lower panel shows the relative differences between
the results from the model and the simulation. \textbf{The gray areas
indicate the $\pm10$\% range.}

\textbf{Similarly, }in Fig.\ref{fig:4bins}, we present the 4-bin
tomographic results where the redshift binning is defined so that
the number density of source galaxies in each bin is the same with
10 arcmin$^{-2}$. It is seen that our model predictions agree with
simulation results very well both in the 2-D case and in the tomographic
peak calculations.

With photometric redshift errors, to generate the tomographic convergence
maps based on the photo-z binning, we need to first calculate the
corresponding true redshift distribution for a particular photo-z
bin. Following \citet{2006ApJ...636...21M}, the photometric redshift
($z^{{\rm ph}}$) distribution given a true redshift $z$ is modeled
by

\begin{equation}
p\left(z^{\mathrm{ph}}|z;z_{{\rm bias}},\sigma_{{\rm ph}}\right)=\frac{1}{\sigma_{{\rm ph}}(1+z)\sqrt{2\pi}}\exp\left[-\frac{\left(z-z^{\mathrm{ph}}-z_{\mathrm{bias}}\times(1+z)\right)^{2}}{2\left(\sigma_{{\rm ph}}\times(1+z)\right){}^{2}}\right],\label{eq:phz_errormodel-1}
\end{equation}
where the redshift-dependent bias and the scatter are given by $z_{{\rm bias}}(1+z)$
and $\sigma_{{\rm ph}}(1+z)$ with $z_{{\rm bias}}$ and $\sigma_{{\rm ph}}$
being constants. With this model, the true redshift distribution in
the photo-z bin of $\left[z_{{\rm ph}}^{(i)},z_{{\rm ph}}^{(i+1)}\right]$
is

\begin{equation}
\ensuremath{p_{i}^{\mathrm{true}}\left(z|z_{\mathrm{ph}}^{(i)}<z_{\mathrm{ph}}<z_{\mathrm{ph}}^{(i+1)}\right)=\int_{z_{\mathrm{ph}}^{(i)}}^{z_{\mathrm{ph}}^{(i+1)}}\mathrm{d}z^{\mathrm{ph}}p(z)\times p\left(z^{\mathrm{ph}}|z;z_{{\rm bias}},\sigma_{{\rm ph}}\right)}
\end{equation}
which gives

\begin{equation}
p_{i}^{\mathrm{true}}(z|z_{\mathrm{ph}}^{(i)}<z_{\mathrm{ph}}<z_{\mathrm{ph}}^{(i+1)})=\frac{1}{2}p(z)\left[{\rm erf}\left(\frac{z_{\mathrm{ph}}^{(i+1)}-z+z_{\mathrm{bias}}(1+z)}{\sqrt{2}\sigma_{z}(1+z)}\right)-{\rm erf}\left(\frac{z_{\mathrm{ph}}^{(i)}-z+z_{\mathrm{bias}}(1+z)}{\sqrt{2}\sigma_{z}(1+z)}\right)\right].\label{eq:n_true_perbin-1}
\end{equation}

Given a set of photo-z error parameters $z_{{\rm bias}}$ and $\sigma_{{\rm ph}}$,
Eq.\eqref{eq:n_true_perbin-1} is used in Eq.\eqref{eq:kappa_mock}
to generate the tomographic convergence map for the photo-z bin of
$z_{{\rm ph}}\in\left[z_{{\rm ph}}^{(i)},z_{{\rm ph}}^{(i+1)}\right]$.
The normalization factor $A$ is calculated correspondingly. We will
describe more in Sec.5 to analyze the impact of photo-z errors in
cosmological studies.

\section{THE DEPENDENCE OF COSMOLOGICAL GAINS ON THE NUMBER OF REDSHIFT BINS}

\begin{figure}[b]
\begin{centering}
\includegraphics[scale=0.3]{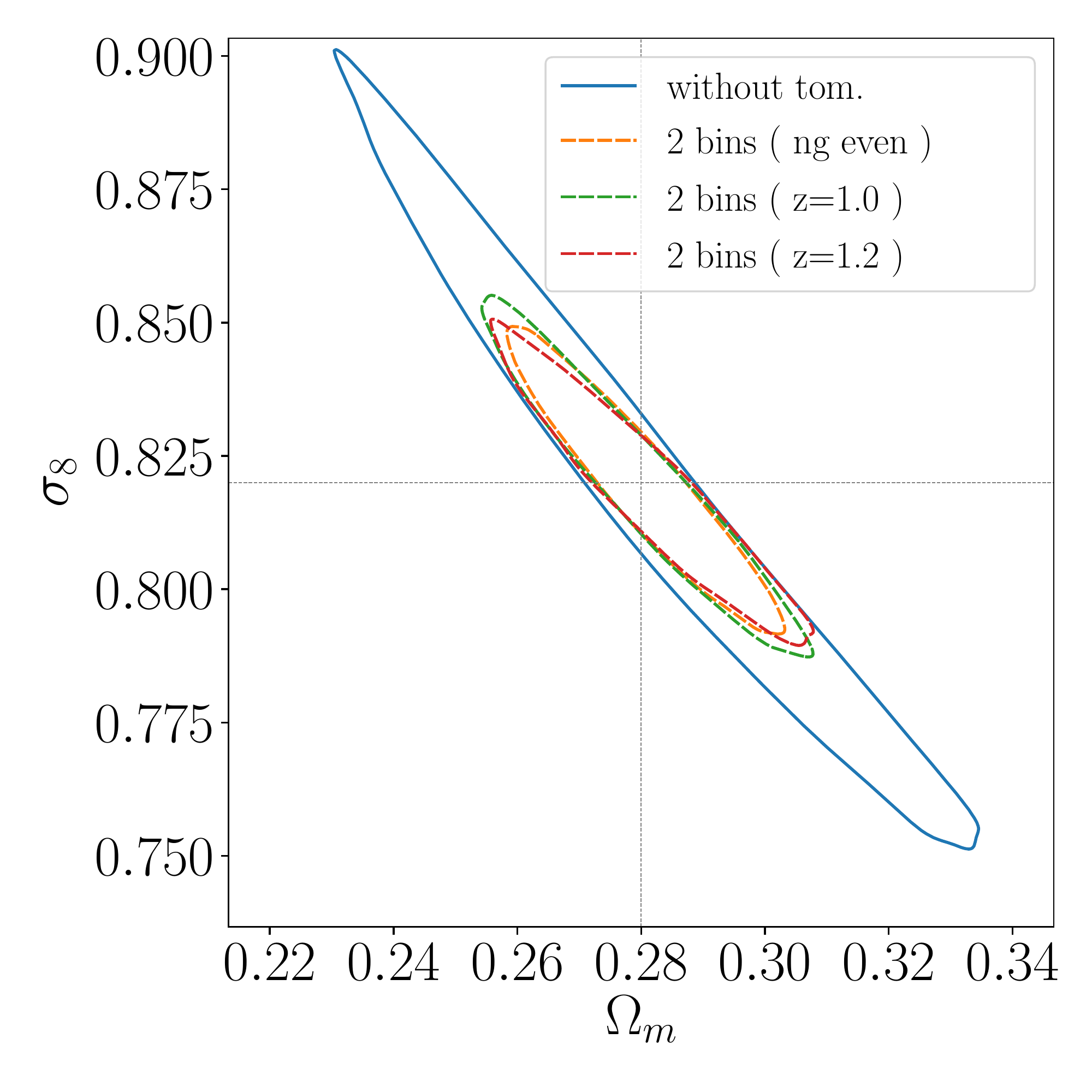}\includegraphics[scale=0.3]{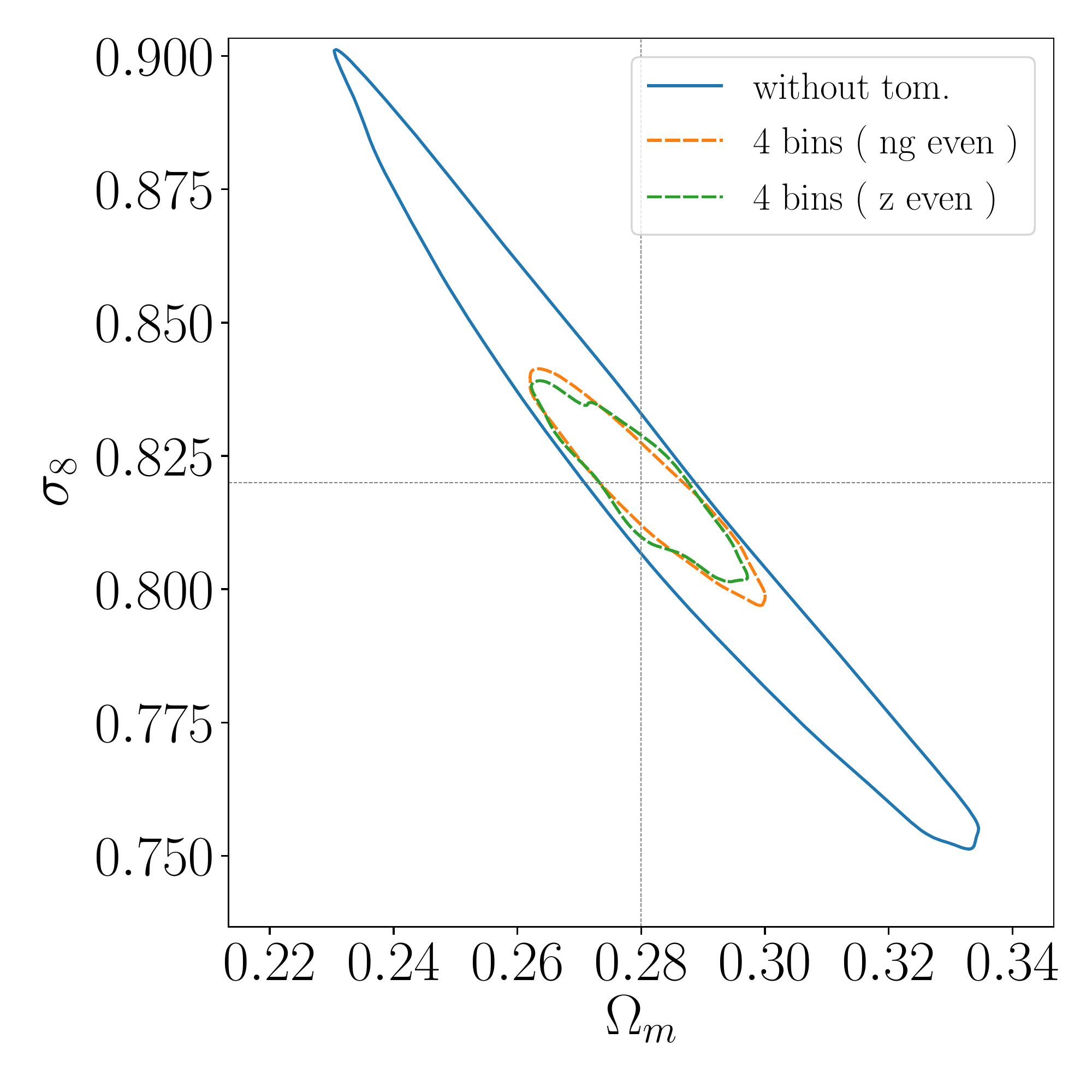}\includegraphics[scale=0.3]{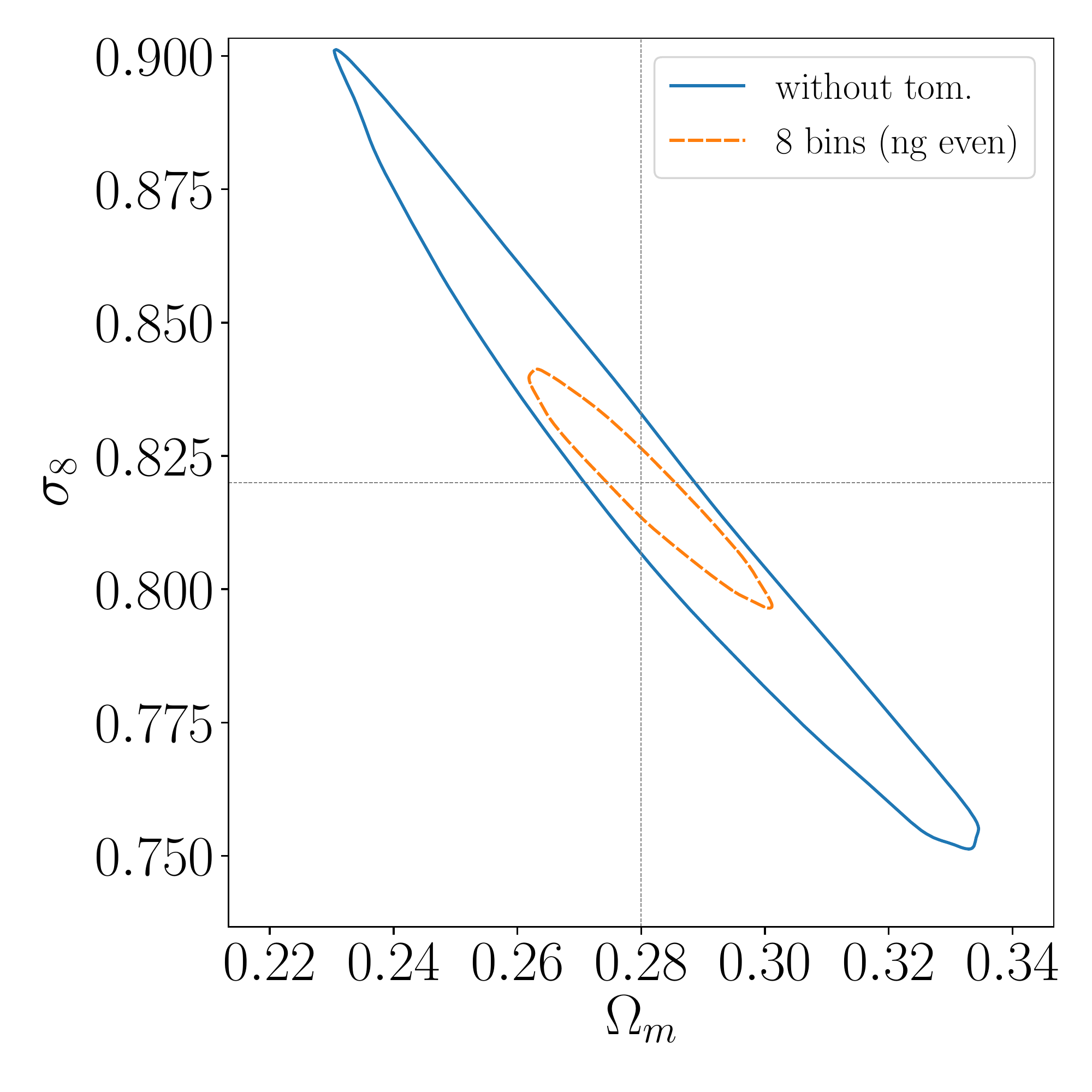}
\par\end{centering}
\figcaption{\label{fig:tom_MCMC}The constraints of 1-$\sigma$ contours in different
cases. The left, middle and right panels present the 2-bin, 4-bin
and 8-bin results with the binning schemes shown in the plots. }
\end{figure}

In this section, we investigate the optimal redshift binning in tomographic
high WL peak studies with respect to the cosmological information
enhancement. Here we consider the ideal case without photo-z errors.
Intuitively, more redshift bins can provide more information about
the evolution of large-scale structures, and thus increase the gains.
However, given a survey, the number of galaxies per bin decreases
with the increase of bin numbers, and the statistical uncertainties
increase correspondingly. Therefore there should exist an optimal
range of bins for cosmological studies. For tomographic 2PCF analyses,
it is found that typically 5-10 bins give best cosmological constraints,
and a further increase of the number of bins cannot lead to significant
improvements \citep{2012JCAP...04..034H,2006ApJ...636...21M}.

Here we study how the number of redshift bins $N_{{\rm bin}}$ affects
the cosmological constraints from tomographic high WL peak analyses.
For that, we consider $N_{{\rm bin}}=0,2,4$ and 8, respectively.
For each $N_{{\rm bin}}$, we construct a set of tomographic convergence
maps as described in Sec. 3. From these maps, we identify peaks with
$\nu\gtrsim4$ and then group them to obtain the peak abundances in
different bins with the width of $\Delta\nu=0.5$. We then calculate
the covariance of the peak abundances for an area of $3.5\times3.5\ {\rm deg^{2}}$
(excluding the outer 70 pixels in each side) from the 96 maps, and
scale the covariance to the considered area.

It is known that the $\Omega_{m}-\sigma_{8}$ constraints from WL
analyses generally show a banana shape. When the survey area is relatively
small, this deviates from an ellipse considerably, indicating that
the Fisher matrix analyses can lead to some errors \citep{PhysRevD.77.042001,2006JCAP...10..013P,2014MNRAS.441.1831S,2016MNRAS.456.1645S,BRINCKMANN2019100260}.
Thus in this section to perform cosmological studies with different
$N_{{\rm bin}}$, we do not use the Fisher matrix forecast. Instead,
we carry out more general MCMC fitting. In the next section to investigate
the impact of photo-z errors, we focus on a large survey area of 15000
${\rm deg^{2}}$. For that, the expected statistical errors are small.
We thus apply the Fisher analyses there, which should be applicable
to a high degree and are more efficient than MCMC fitting when photo-z
error parameters are also included in the study.

\textbf{We adopt the same likelihood analysis procedures as in \citealt{2018ApJ...857..112Y}.
The $\chi^{2}$ is defined as:
\begin{equation}
\chi^{2}\equiv-2\ln\mathcal{L}=\mathbf{\Delta^{T}\widehat{\mathbf{C}^{-1}}\Delta},\label{eq:chi2}
\end{equation}
where $\mathbf{\Delta=N-\hat{N}}$ with $\mathbf{N}$ being the mock
data vector of WL peak counts of different bins and $\mathbf{\hat{N}}$
being the theoretical predictions for these bins. For the covariance,
in this paper, we first calculate them directly from simulated maps
for an area of $3.5\times3.5\deg^{2}$, denoted as $S_{0}$. For that,
for each of the 96 maps, we generate 20 shape noises with different
random seeds. Thus we totally have $20\times96$ maps. From them,
we compute the covariance matrix $\mathbf{C_{0}}$ of peaks. For the
covariance $\mathbf{C}$ of a large area $S$, we use the scaling
relation of $\mathbf{C}=(S/S_{0})\mathbf{C_{0}}$. It is noted that
this scaling does not include contributions to the covariance from
scales larger than $S_{0}$. Thus it can lead to a slight underestimate
of the covariance for an area of $S$ \citep{2010PhRvD..81d3519K,2015MNRAS.450.2888L}.
For more precise analyses, we need to run many large simulations to
produce many maps matching the considered survey area to calculate
the covariance matrix. This can be difficult. Certain approximated
and fast simulation methods have been proposed \citep{2018JCAP...10..051F}.
The full analyses of the covariance calculations are beyond the scope
of the current paper, and will be studied in our future investigations.}

\textbf{To calculate the inverse covariance, we adopt the unbiased
estimator used in \citet{2007A&A...464..399H}, which is given by}

\textbf{
\begin{equation}
\widehat{\mathbf{C}^{-1}}=\frac{R-N_{\mathrm{bin}}-2}{R-1}\mathbf{C}^{-1}
\end{equation}
where $R=20\times96$ and $N_{{\rm bin}}$ is the number of bins
of WL peak counts used in deriving cosmological constraints, and $\mathbf{C^{-1}}$
is the normal inverse of $\mathbf{C}$. }

From Fig.\ref{fig:wihtout_tom} and Fig.\ref{fig:4bins}, we see that
our theoretical model for high WL peak abundances works very well.
Thus for clarity, in this section, we perform cosmological parameter
forecasts for different values of $N_{{\rm bin}}$ with mock observational
data generated directly from our model calculations and the covariance
from simulations. Here we consider the survey area of $\sim876\,{\rm deg^{2}}$,
the same as the effective area of our simulations. The improvement
on the cosmological information gains from tomographic peak analyses
is evaluated by comparing the derived constraints with that from the
2-D peak analyses.

In Fig.\ref{fig:tom_MCMC}, we show the 1-$\sigma$ confidence regions
of $\Omega_{m}-\sigma_{8}$ for different cases. In all the panels,
the blue contour is from the 2-D peak analyses. The left panel shows
the results of $N_{{\rm bin}}=2$, where three different binning methods
are considered. The orange one is from dividing galaxies into two
equal-number-density bins, and the green and red ones are using $z=1$
and $z=1.2$ as a dividing point, respectively. The middle panel is
for the results of $N_{{\rm bin}}=4$ with the orange contour from
the equal-number-density binning and the green one from the equal-z-interval
binning, respectively. The right panel is for $N_{{\rm bin}}=8$,
and only the result from the equal-number-density binning is shown.
By comparing with the blue contour in each panel, we see very clearly
that tomographic peak analyses can indeed enhance the cosmological
information significantly. The improvement from $N_{{\rm bin}}=2$
to $N_{{\rm bin}}=4$ is apparent. However, for $N_{{\rm bin}}=8$,
the constraint is nearly the same as that of $N_{{\rm bin}}=4$.

\begin{figure}[H]
\begin{centering}
\includegraphics[scale=0.4]{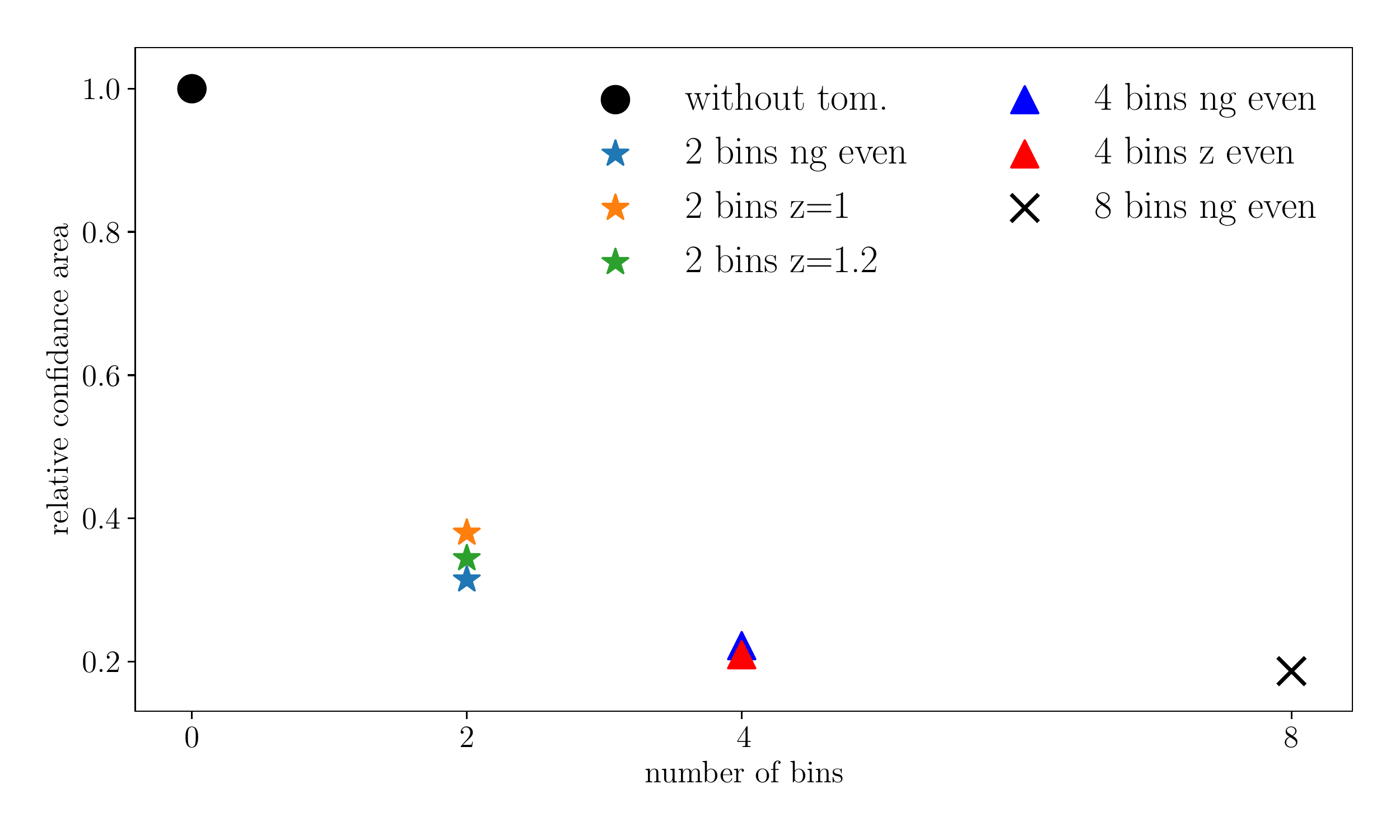}
\par\end{centering}
\begin{centering}
\includegraphics[scale=0.4]{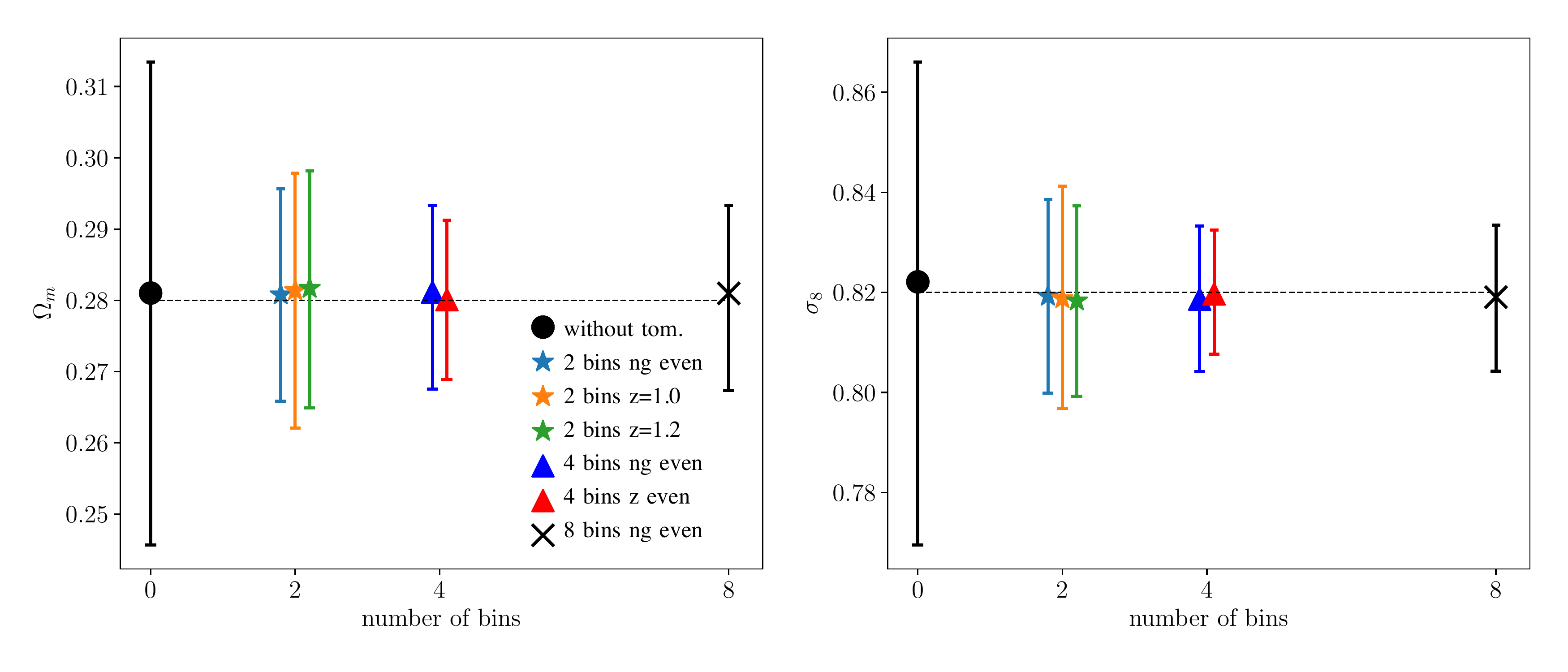}
\par\end{centering}
\figcaption{\label{fig:areas_tom}Measurements of cosmological information gains.
\textit{Upper panel:} Ratio of 1-$\sigma$ confidence region with
respect to non-tomographic case. \textit{Lower panel:} The marginalized
1-$\sigma$ confidence intervals for $\Omega_{m}$ and $\sigma_{8}$
for different tomographic strategies. The error bars are 1-$\sigma$
errors extracted from MCMC.}
\end{figure}

To be more quantitative, we calculate the area within 1-$\sigma$
confidence region in different cases, and the results relative to
the 2-D case are shown in the upper panel of Fig.\ref{fig:areas_tom}.
It is seen that from 2-D to $N_{{\rm bin}}=2$, the constraining area
is reduced by about 2.5 times, which is about 40\% of that of the
2-D case without tomography. With $N_{{\rm bin}}=4$, the area further
decreases by about a factor of two. Further to $N_{{\rm bin}}=8,$
the improvement is not significant. In the lower panels of Fig.\ref{fig:areas_tom},
we show the comparisons of the corresponding 1-D constraints for $\Omega_{m}$
(left) and $\sigma_{8}$ (right). The trend is the same as that shown
in the upper panel.

We thus conclude that considering high WL peaks, with the source redshift
distribution similar to LSST, tomographic peak analyses with $N_{{\rm bin}}\sim4$
can give optimal cosmological constraints, which can improve the 1-$\sigma$
confidence area of $(\Omega_{m},\sigma_{8})$ by a factor of 5 with
respect to the 2-D peak analyses without redshift binning. To increase
the number of bins further cannot lead to significantly better constraints.

\section{IMPACT OF PHOTOMETRIC REDSHIFT ACCURACY ON TOMOGRAPHIC PEAK ANALYSES}

In Sec.4, we studied the cosmological information improvement from
tomographic peak analyses in the ideal case that the galaxy redshifts
are perfectly known. In practice, however, for WL cosmological studies,
we need to observe a large number of far-way source galaxies that
are typically faint. It is therefore very difficult to obtain spectroscopic
redshifts (spec-z) for all the galaxies. The feasible way is to estimate
their photo-z by multi-band observations. The accuracy of photo-z
depends on observations, such as the overall wavelength coverage,
the central position and band-width of filters, photometric accuracy,
etc., as well as the methodology for photo-z estimates \citep{2010A&A...523A..31H,2019NatAs...3..212S}.

The photo-z errors are often characterized by the bias, the scatter,
and the fraction of outliers, estimated using a subsample of galaxies
with known spec-z \citep{2019NatAs...3..212S}. It is noted that not
only these information, but also their uncertainties, namely, errors
on errors, are important in tomographic WL studies. Their effects
on tomographic 2PCF analyses and the corresponding error propagation
to cosmological studies have been investigated extensively \citep{2006ApJ...636...21M,2006MNRAS.366..101H,2008ApJ...682...39M,2008MNRAS.391..228A,2009ApJ...699..958S,2010MNRAS.401.1399B,2010ApJ...720.1351H,2012JCAP...04..034H,2017JCAP...10..056Y}.
For WL peak studies, some of the photo-z effects are explored using
numerical simulations \citep{2016PhRvD..94f3534P,2018arXiv181012312A}.

In this section, the impacts of photo-z errors on tomographic high
WL peak studies are analyzed. We investigate how the photo-z error
parameters can be constrained simultaneously with cosmological parameters,
and the corresponding degradations of the cosmological constraints.
We then study how the knowledge about photo-z errors on errors can
improve the degradations. Here we consider the survey area of 15000
${\rm deg}^{2}$, and adopt the Fisher forecast approach to efficiently
explore the multi-dimensional parameter space.

\subsection{Photometric redshift errors}

For easy-read purposes, we list the photo-z relevant formulae again
here. The photo-z distribution given a true redshift $z$ is taken
to be Gaussian, given by 

\begin{equation}
p\left(z^{\mathrm{ph}}|z;\sigma_{{\rm ph}},z_{{\rm bias}}\right)=\frac{1}{\sigma_{{\rm ph}}(1+z)\sqrt{2\pi}}\exp\left[-\frac{\left(z-z^{\mathrm{ph}}-z_{\mathrm{bias}}(1+z)\right)^{2}}{2\left(\sigma_{{\rm ph}}(1+z)\right){}^{2}}\right],\label{eq:phz_errormodel}
\end{equation}
where $\sigma_{{\rm ph}}\times(1+z)$ and $z_{{\rm bias}}\times(1+z)$
are the redshift-dependent scatter (precision) and bias with $\sigma_{{\rm ph}}$
and $z_{{\rm bias}}$ being constants to be constrained and analyzed
from tomographic peak abundances. Then the true redshift distribution
given the photometric redshift interval $\left[z_{{\rm ph}}^{(i)},z_{{\rm ph}}^{(i+1)}\right]$
can be calculated to be

\begin{equation}
p_{i}^{\mathrm{true}}(z|z_{\mathrm{ph}}^{(i)}<z_{\mathrm{ph}}<z_{\mathrm{ph}}^{(i+1)})=\frac{1}{2}p(z)\left[{\rm erf}\left(\frac{z_{\mathrm{ph}}^{(i+1)}-z+z_{\mathrm{bias}}(1+z)}{\sqrt{2}\sigma_{z}(1+z)}\right)-{\rm erf}\left(\frac{z_{\mathrm{ph}}^{(i)}-z+z_{\mathrm{bias}}(1+z)}{\sqrt{2}\sigma_{z}(1+z)}\right)\right].\label{eq:n_true_perbin}
\end{equation}

In the Fisher analyses here, we adopt the photo-z error parameters
as that required by LSST-like surveys, with the fiducial vales of
$z_{{\rm bias}}=0.003$ and $\sigma_{{\rm ph}}=0.02,$ respectively.
The overall true redshift distribution $p(z)$ is given by Eq.\eqref{eq:prob_z_LSST}.
Meanwhile we scale the covariance calculated from our mock simulations
to the survey area of 15000 ${\rm deg^{2}}$. Note that in this study,
we do not consider catastrophic photo-z errors with large deviations
from the true redshifts that cannot be described by Eq.\eqref{eq:phz_errormodel}
\citep{2009ApJ...699..958S}.

The following two cases are considered including the photo-z errors:
\begin{enumerate}
\item 2-bins-tomography divided by photometric redshift $z^{{\rm ph}}$=1.0;
\item 4-bins-tomography divided by even source galaxies number density adopt
from photometric redshift.
\end{enumerate}
~

In our model calculations, we include the photo-z error parameters
by adopting the distribution of Eq.\eqref{eq:n_true_perbin}. The
likelihood is then updated to include four free parameters $\mathcal{L}(\Omega_{m},\sigma_{8},z_{{\rm bias}},\sigma_{{\rm ph}})$
to be constrained simultaneously.

\subsection{Fisher analyses and error propagation}

In the Fisher approximation, the error propagation from data to cosmological
parameters can be estimated by the Fisher matrix given by: \citep{2006ApJ...636...21M,2016Entrp..18..236H}

\begin{equation}
\mathbf{F}_{\alpha\beta}=\left\langle -\frac{\partial^{2}\ln\mathcal{L}(\Omega_{m},\sigma_{8},z_{{\rm bias}},\sigma_{{\rm ph}})}{\partial_{\alpha}\partial_{\beta}}\right\rangle ,\label{eq:def_fisher_matrix}
\end{equation}
where $\alpha,\beta=1,2,3,4$ are the parameters to be constrained.
Assuming Gaussian priors for both $z_{{\rm bias}}$ and $\sigma_{{\rm ph}}$,
their prior matrix can be written as

\begin{equation}
\mathbf{P}=\left(\begin{array}{cccc}
0 & 0 & 0 & 0\\
0 & 0 & 0 & 0\\
0 & 0 & \ensuremath{\left[\sigma^{\text{prior}}\left(z_{\text{bias}}\right)\right]^{-2}} & 0\\
0 & 0 & 0 & \ensuremath{\left[\sigma^{{\rm prior}}\left(\sigma_{\mathrm{ph}}\right)\right]^{-2}}
\end{array}\right),\label{eq:fisher_prior_matrix}
\end{equation}
where the $\sigma^{\text{prior}}\left(z_{\text{bias}}\right)$ and
$\sigma^{{\rm prior}}\left(\sigma_{\mathrm{ph}}\right)$ are the standard
dispersions of the corresponding Gaussian priors. According to the
Bayesian theory, the  Fisher matrix of posterior  is:

\begin{equation}
\mathbf{F}^{{\rm post}}=\mathbf{F}+\mathbf{P}.\label{eq:fisher_post}
\end{equation}
This is used to forecast the parameter constraints by computing its
inverse matrix as follows

\begin{equation}
\mathbf{C}^{{\rm post}}=\left(\mathbf{F}^{{\rm post}}\right){}^{-1}.\label{eq:cov_by_fisher_post}
\end{equation}

To calculate the Fisher matrix Eq.\eqref{eq:def_fisher_matrix}, because
we have our theoretical model for tomographic high peak abundances,
in principle, we can compute the derivatives directly from the model.
However, such estimates can suffer from numerical instabilities. To
avoid these, we run MCMC fitting for the case without any priors on
the photo-z error parameters. From the converged sampling chains,
we obtain an estimate for the corresponding covariance by

\begin{equation}
\mathbf{C}_{\alpha\beta}^{{\rm MCMC}}=\frac{1}{N-1}\sum_{i=1}^{N}\left(\theta_{\alpha}^{(i)}-\theta_{\alpha}^{*}\right)\left(\theta_{\beta}^{(i)}-\theta_{\beta}^{*}\right),
\end{equation}
where $\boldsymbol{\theta}^{(i)}$ denotes the point in the four-dimensional
parameter space at the $i$-th sampling step and $N$ is the number
of total steps in MCMC chains and $\theta_{\alpha}^{*}(\alpha=1,2,3,4)$
are the mean values obtained from the MCMC chains. Its inverse matrix
gives rise to an estimation of the Fisher matrix corresponding to
Eq.\eqref{eq:def_fisher_matrix} without the photo-z priors, which
is given by

\begin{equation}
\mathbf{F}_{\alpha\beta}=\left(\mathbf{C}_{\alpha\beta}^{{\rm MCMC}}\right){}^{-1}.\label{eq:fisher_fromMCMC}
\end{equation}
This is then used in Eq.\eqref{eq:fisher_prior_matrix} to Eq.\eqref{eq:cov_by_fisher_post}
to evaluate how the prior knowledge on photo-z error parameters can
affect the cosmological parameter constraints.

\subsection{High peak tomography with photometric redshift errors}

\begin{figure}[b]
\begin{centering}
\includegraphics[scale=0.5]{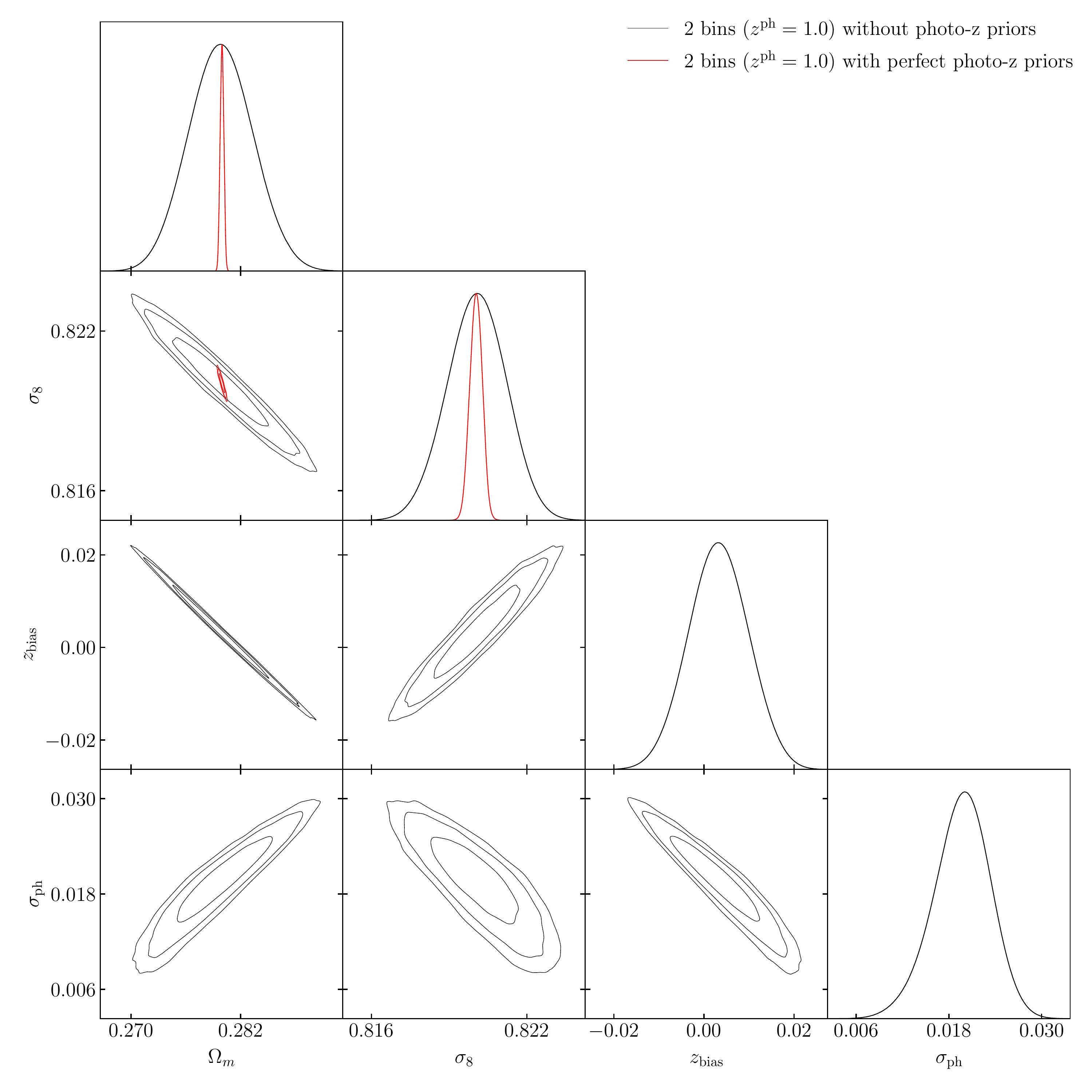}
\par\end{centering}
\figcaption{\label{fig:phz_mcmc-left}Parameter forecasts for 2-bin peak tomography
(divided by $z^{{\rm ph}}=1.0$). The black curves are from the model
with 4 free parameters $\{\Omega_{m},\sigma_{8},z_{{\rm bias}},\sigma_{{\rm ph}}\}$
while the red ones are for the case with perfectly known photo-z parameters.}
\end{figure}

We first study the 2-bin case with the dividing redshift set to be
$z^{{\rm ph}}=1.0$. In Fig.\ref{fig:phz_mcmc-left}, we show the
constraining results without priors on photo-z error parameters obtained
from MCMC fitting (black contours). For comparison, the cosmological
parameter constraints with perfectly known photo-z errors, i.e., fixing
their values to the fiducial ones with $z_{{\rm bias}}=0.003$ and
$\sigma_{{\rm ph}}=0.02$, are also shown in the figure (in red).
It is seen from the black contours that, in the 2-bin case, there
are strong degeneracies between the photo-z error parameters and the
cosmological parameters. This leads to a severe degradation in cosmological
parameter constraints if no prior knowledge on the photo-z error parameters
is known.

\begin{figure}[H]
\noindent \begin{centering}
\includegraphics[scale=0.4]{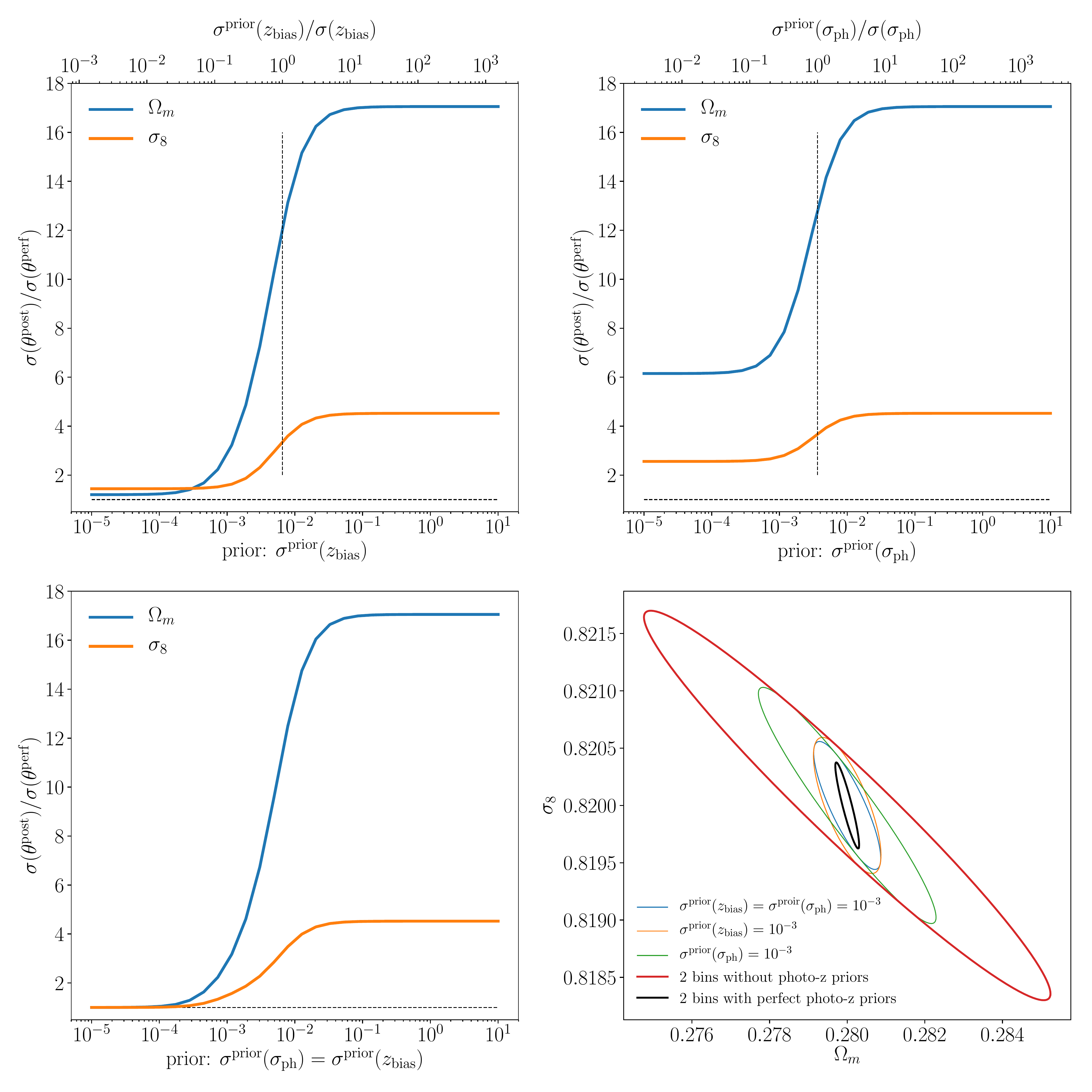}
\par\end{centering}
\centering{}\figcaption{\label{fig:phz_degradation-upper} Degradations for the 2-bin case.
The upper panels show the degradations with respect to $\sigma^{{\rm prior}}(z_{{\rm bias}})$
(left) and $\sigma^{{\rm prior}}(\sigma_{{\rm ph}})$ (right), respectively.
For each panel, there is no prior applied to the other photo-z parameter.
The bottom left panel is the degradation with priors on both parameters
assuming $\sigma^{{\rm prior}}(z_{{\rm bias}})=\sigma^{{\rm prior}}(\sigma_{{\rm ph}})$.
The bottom right is the 1-$\sigma$ constraining contours under different
priors. }
\end{figure}

We then add different priors to the photo-z parameters to study the
improvements in the cosmological parameter constraints. The results
are shown in Fig.\ref{fig:phz_degradation-upper}. The first and the
second panels are for the results adding the prior only to $z_{{\rm bias}}$
or $\sigma{\rm _{ph}}$ , respectively, leaving the other one fully
free without prior. The vertical axes are the degradation factor with
respect to the case with perfectly known $z_{{\rm bias}}=0.003$ and
$\sigma_{{\rm ph}}=0.02$. We also show the degradation by assuming
equal priors on $z_{{\rm bias}}$ and $\sigma{\rm _{ph}}$, i.e.,
$\ensuremath{\sigma^{{\rm prior}}\left(z_{\text{ bias}}\right)=\sigma^{\text{prior}}\left(\sigma_{\mathrm{ph}}\right)}$,
in the lower left panel. The lower right panel is for the cosmological
constraints of $\{\Omega_{m},\sigma_{8}\}$ with a few examples of
different priors on the photo-z parameters. From the plots, we see
clearly that the error on photo-z bias parameter $z_{{\rm bias}}$
affects more importantly in the cosmological constraints than that
of scatter parameter $\sigma_{{\rm ph}}$. Without any prior on $\sigma_{{\rm ph}}$,
the degradations on the cosmological parameters are $\lesssim1.5$
if the prior on $z_{{\rm bias}}$ can reach the level of $\ensuremath{\sigma^{{\rm prior}}\left(z_{\text{ bias }}\right)<10^{-4}}$.
However, in the case that $z_{{\rm bias}}$ is completely free from
any prior, the best degradations we can get for $\Omega_{m}$ and
$\sigma_{8}$ are $\sim6$ and $\sim2.5$, respectively, even with
the prior on $\sigma_{{\rm ph}}$ reaching zero. The vertical dotted
lines in the first and the second panels indicate the location where
the prior is equal to the constraint shown by the corresponding black
line in Fig.\ref{fig:phz_mcmc-left} obtained solely from the 2-bin
tomographic peak abundances without any priors on photo-z parameters.

\begin{figure}[H]
\begin{centering}
\includegraphics[scale=0.5]{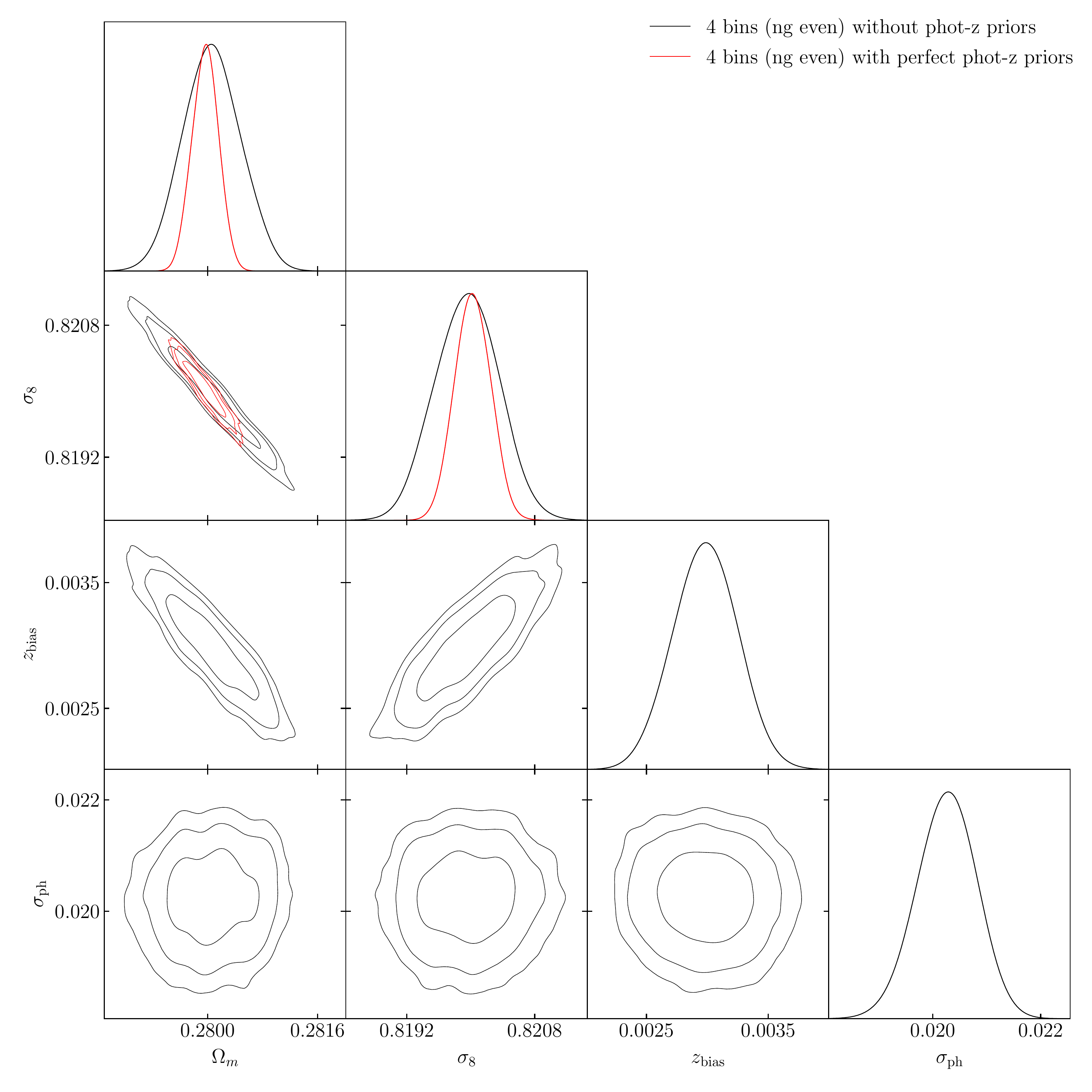}
\par\end{centering}
\figcaption{\label{fig:phz_mcmc-right} Similar to Fig.\ref{fig:phz_mcmc-left},
but for the 4-bin case with an equal number density of source galaxies
in each bin. }
\end{figure}

\begin{figure}[H]
\noindent \begin{centering}
\includegraphics[scale=0.4]{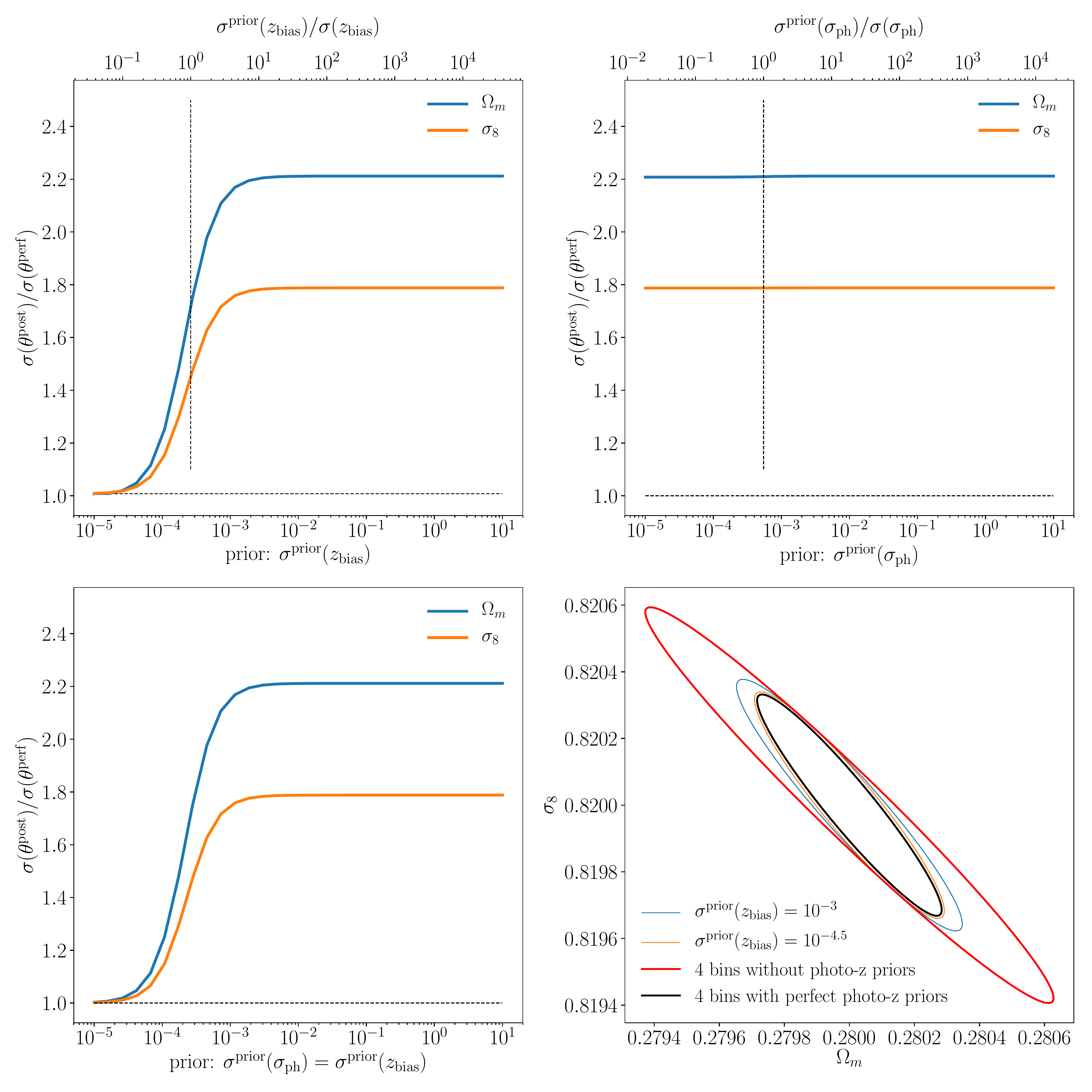}
\par\end{centering}
\centering{}\figcaption{\label{fig:phz_degradation-lower} Similar to Fig.\ref{fig:phz_degradation-upper}
but for the 4-bin case.}
\end{figure}

Now we show the results for the 4-bin case with equal number density
of source galaxies in each bin. The parameter forecasts for models
with free parameters of $\{\Omega_{m},\sigma_{8},z_{{\rm bias}},\sigma_{{\rm ph}}\}$
(black) and the case with fixed $z_{{\rm bias}}=0.003$ and $\sigma_{{\rm ph}}=0.02$
(red) are shown Fig.\ref{fig:phz_mcmc-right}. Similar to the 2-bin
case, without any priors on the photo-z parameters, the information
of cosmological parameters is also degraded, but the level is smaller
than that of the 2-bin case. We also see that with 4-bin tomographic
peak abundances, we can constrain $z_{{\rm bias}}$ and $\sigma_{{\rm ph}}$
much better than that of the 2-bin case. There is still an apparent
degeneracy between the cosmological parameters and $z_{{\rm bias}}$,
but their correlations with $\sigma_{{\rm ph}}$ are insignificant.
The degradation curves and 1-$\sigma$ contours for different priors
on the photo-z parameters are shown in Fig.\ref{fig:phz_degradation-lower}.
It is seen that adding prior on $\sigma_{{\rm ph}}$ alone has nearly
no impact on the degradation on cosmological parameter constraints
(upper right panel). On the other hand, if we control the prior on
$z_{{\rm bias}}$ to the precision of $\ensuremath{\sigma^{{\rm prior}}\left(z_{\text{bias}}\right)\sim10^{-5}}$
without any prior on $\sigma_{{\rm ph}}$, the degradation factors
can be close to unit (upper left panel). This is consistent with the
degeneracy behaviors shown in Fig.\ref{fig:phz_mcmc-right}.

Comparing the degradation curves of the above 2 cases (the upper left
and the upper right panels in Fig.\ref{fig:phz_degradation-upper}
and Fig.\ref{fig:phz_degradation-lower}), we find that the prior
on the bias parameter is more important for improving the cosmological
information gain. From the lower left panel of Fig.\ref{fig:phz_degradation-upper}
and Fig.\ref{fig:phz_degradation-lower}, it is seen that the results
are about the same as those applying priors only to $z_{{\rm bias}}$,
which demonstrate again the importance of knowing $z_{{\rm bias}}$
accurately.

To see the dependence of the degradation factors on the priors of
$z_{{\rm bias}}$ and $\sigma_{{\rm ph}}$ more clearly, we use the
Sherman\textendash Morrison formula, which shows that the inverse
of $\mathbf{F}$ and $\mathbf{F}^{{\rm post}}$ defined in Eq.\eqref{eq:fisher_post}
are related by:

\begin{equation}
\left(\mathbf{F}^{{\rm post}}\right)^{-1}\equiv\left(\mathbf{F}+\mathbf{P}\right)^{-1}=\mathbf{F}^{-1}-\frac{\mathbf{F^{-1}PF^{-1}}}{1+{\rm Tr}(\mathbf{PF^{-1}})}.\label{eq:sherman-morrison_eq}
\end{equation}
If only the prior on parameter $\beta$ is considered, the posterior
error of the other parameters can be computed from the diagonal elements
of left hand side of Eq.\eqref{eq:sherman-morrison_eq} (For details
please see Eq.(3) \textendash{} Eq.(10) in \citealp{2016MNRAS.457.1490A}.),
which is

\begin{equation}
\ensuremath{\left(\sigma_{\alpha}^{\mathrm{post}}\right)^{2}=\sigma_{\alpha}^{2}-\frac{\rho_{\alpha\beta}^{2}\sigma_{\alpha}^{2}}{1+\left(\sigma_{\beta}^{\mathrm{prior}}\right)^{2}/\sigma_{\beta}^{2}}},\label{eq:shermann-morrison_eq_scalar}
\end{equation}
where $\sigma_{\alpha}$ and $\sigma_{\beta}$ are the standard dispersion
obtained without any priors, and $\rho_{\alpha\beta}$ is the correlation
coefficient between the two parameters (without any priors added as
well). The quantity $\sigma_{\alpha}^{\mathrm{post}}$ is the dispersion
from the posterior distribution defined in Eq.\eqref{eq:fisher_post}
including the a prior with a dispersion of $\sigma_{\beta}^{\mathrm{prior}}$
on parameter $\beta$.

We can see that the effect of the prior of $\beta$ on the constraint
of the parameter $\alpha$ depends on their correlation $\rho_{\alpha\beta}$,
the larger the correlation, the stronger the effect. The correlation
coefficients between different parameters in the 2-bin and 4-bin cases
are shown in Table.\ref{tab:correlation-coefficients}. It is seen
that in the 4-bin case, the correlations between $\sigma{\rm _{ph}}$
and the other parameters are very small, thus its prior has almost
no effect on the degradation factors of $\Omega_{m}$ and $\sigma_{8}$
as seen in the upper right panel of Fig.\ref{fig:phz_degradation-lower}.
On the other hand, the correlations between $z_{{\rm bias}}$ and
the cosmological parameters are large, therefore the prior on $z_{{\rm bias}}$
is much more important in improving the degradations than that of
$\sigma_{{\rm ph}}$.

\begin{table}[H]
\caption{\label{tab:correlation-coefficients}The correlation coefficients
for 2-bins-tomography and 4-bins-tomography (in bold).}

\begin{singlespace}
\noindent \begin{centering}
\begin{tabular}{cccc|c}
$\Omega_{m}$ & $\sigma_{8}$ & $z_{{\rm bias}}$ & $\sigma_{{\rm ph}}$ & $\rho_{\alpha\beta}$\tabularnewline
\hline 
\multirow{2}{*}{1} & -0.966 & -0.997 & 0.944 & \multirow{2}{*}{$\Omega_{m}$}\tabularnewline
 & \textbf{-0.974} & \textbf{-0.891} & \textbf{0.060} & \tabularnewline
\cline{1-4} \cline{2-4} \cline{3-4} \cline{4-4} 
\multirow{2}{*}{} & \multirow{2}{*}{1} & 0.947 & -0.824 & \multirow{2}{*}{$\sigma_{8}$}\tabularnewline
 &  & \textbf{0.826} & \textbf{0.031} & \tabularnewline
\cline{2-4} \cline{3-4} \cline{4-4} 
\multirow{2}{*}{} & \multirow{2}{*}{} & \multirow{2}{*}{1} & -0.947 & \multirow{2}{*}{$z_{{\rm bias}}$}\tabularnewline
 &  &  & \textbf{0.045} & \tabularnewline
\cline{3-4} \cline{4-4} 
\multirow{2}{*}{} & \multirow{2}{*}{} & \multirow{2}{*}{} & \multirow{2}{*}{1} & \multirow{2}{*}{$\sigma_{{\rm ph}}$}\tabularnewline
 &  &  &  & \tabularnewline
\end{tabular}
\par\end{centering}
\end{singlespace}
\end{table}

\subsection{Photo-z calibration requirements}

In this study, we assume a Gaussian-like conditional probability of
photo-z given a spec-z, with the bias and the dispersion being $z_{{\rm bias}}\times(1+z)$
and $\sigma{\rm _{ph}}\times(1+z)$, respectively. We analyze the
dependence of the degradation of the cosmological constraints on the
priors of the two error parameters $\{z_{{\rm bias}}$, $\sigma{\rm _{ph}\}}$
and show that the prior knowledge on the bias parameter plays more
important roles in improving the cosmological information gain.

\begin{figure}[H]
\begin{centering}
\includegraphics[scale=0.5]{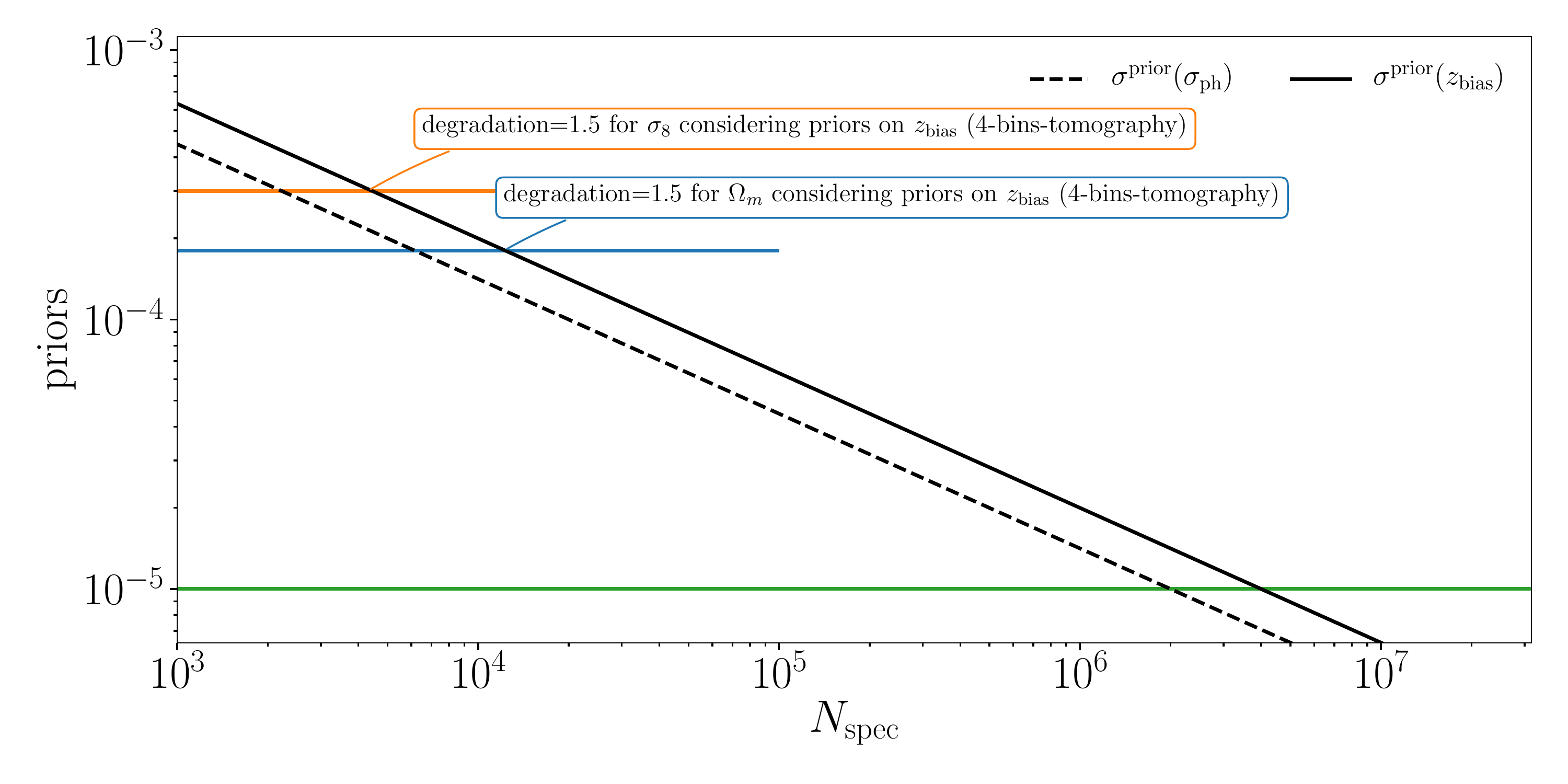}
\par\end{centering}
\figcaption{\label{fig:nspec}The relation between the desired priors on the photo-z
parameters and the number of spec-z required from Eq.\eqref{eq:prior_phz_nspec}.
The horizontal lines indicate the prior requirements for $z_{{\rm bias}}$
of different cases.}
\end{figure}

Observationally, one direct way to calibrate the conditional probability
of photo-z is to use a number of spec-z measurements. Assuming we
have $N_{{\rm spec}}(z)$ spec-z in the a photo-z bin centered at
$z$ and a bin width $\delta z$, and they are fair samples of the
Gaussian distribution. Then the accuracies of the estimates about
$\sigma{\rm _{ph}}\times(1+z)$ and $z_{{\rm bias}}\times(1+z)$ are
\citep{2006ApJ...636...21M}

\begin{equation}
\ensuremath{\begin{aligned}\sigma^{\mathrm{prior}}\left(\sigma_{\mathrm{ph}}\times(1+z)\right) & =\frac{\sigma_{\mathrm{ph}}\times(1+z)}{\sqrt{2N_{\mathrm{spec}}}};\\
\sigma^{\mathrm{prior}}\left(z_{\mathrm{bias}}\times(1+z)\right) & =\frac{\sigma_{\mathrm{ph}}\times(1+z)}{\sqrt{N_{\mathrm{spec}}}}.
\end{aligned}
}\label{eq:prior_phz_nspec}
\end{equation}
It is noted that the $(1+z)$ factor occurs in both sides of the equations,
and thus can be canceled out. They give an estimate about the required
number $N_{{\rm spec}}$ in each bin given a desired prior on the
photo-z errors. The dashed and solid black lines in Fig.\ref{fig:nspec}
shows the above relations taking $\sigma_{{\rm ph}}=0.02$.

For the 4-bin tomographic peak studies, with $\ensuremath{\sigma^{\text{prior}}\left(z_{\text{bias}}\right)\sim10^{-5}}$,
there is nearly no degradation in cosmological parameter constraints
comparing to the case with perfectly known photo-z parameters. To
reach this accuracy, $\ensuremath{N_{\mathrm{spec}}\sim5\times10^{6}}$
is needed for each $\delta z$ redshift bin. We note that $\delta z$
here should be smaller than the bins used for tomographic peak analyses
so that the a single Gaussian distribution centered on $z$ is applicable.
With $\delta z=0.1$, and the source redshifts extending to $z\sim3$,
we need totally $\sim1.5\times10^{8}$ spec-z measurements. This can
be a challenge, particularly for high redshift bins. If we allow a
factor of 1.5 degradation in the cosmological parameter constraints,
in the 4-bin case, the photo-z accuracy can be relaxed to $\ensuremath{\sigma^{{\rm prior}}\left(z_{\text{bias}}\right)\sim10^{-4}}$.
The corresponding requirements for spec-z observations is $\ensuremath{N_{\mathrm{spec}}\sim10^{4}}$
in each bin. We caution here that the Gaussian conditional photo-z
distribution with the form of bias and the dispersion considered in
this paper is relatively simple. More generally, the photo-z errors
in different bins can be different, and not follow the $(1+z)$ dependence.
In that case, we need to include more photo-z error parameters, and
the prior requirements for different photo-z bins can be different.
This can affect the requirements on the number of spec-z observations
\citep{2006ApJ...636...21M}.

In addition to the direct spec-z calibration, variety of other photo-z
calibration methods have been discussed in literature \citep{2019NatAs...3..212S}.
The detailed tomographic WL peak studies taking into account complicated
photo-z distributions and careful examinations of the requirements
for different redshift calibration methods are beyond the scope of
the current paper, and will be addressed in our future investigations.

\section{Conclusions}

In this paper, we investigate the potential of tomographic WL peak
abundance studies, and explore the cosmological gains from the peak
tomography in comparison with that of 2-D peak statistics. We concentrate
on high peaks with signal-to-noise ratio $\nu\geq4$, and adopt our
theoretical model to calculate the fiducial peak abundances in different
tomographic bins. At the same time, we carry out ray-tracing simulations
to validate our model, and also to compute the covariance matrix for
tomographic peak statistics.

Considering LSST-like surveys, we find that 4-bin tomographic peak
analyses can lead to about 5 times better constraints for $\Omega_{m}$
and $\sigma_{8}$ than that from 2-D peak abundances in the ideal
case without considering photo-z errors. Taking into account these
errors, we investigate how they can be constrained simultaneously
with cosmological parameters using tomographic peak abundances alone.
Adopting a Gaussian conditional photo-z distribution with the bias
and the dispersion modeled as $z_{{\rm bias}}\times(1+z)$ and $\sigma_{{\rm ph}}\times(1+z)$,
respectively, we show that 4-bin peak tomography itself can constrain
photo-z error parameters to the level of $\sigma(z_{{\rm bias}})\sim3\times10^{-4}$
, and $\sigma(\sigma_{{\rm ph}})\sim6\times10^{-4}$. The corresponding
cosmological constraints for $\Omega_{m}$ and $\sigma_{8}$ are degradated
by a factor of 2.2 and 1.8, respectively, with respect to the case
with perfectly known $z_{{\rm bias}}$ and $\sigma_{{\rm ph}}$. To
limit the degradation to below 1.5, we need to have a prior knowledge
on z bias with the accuracy of $\sim10^{-4}$ . The priors on $\sigma_{{\rm ph}}$
are less important.

\textbf{Intuitively, the more sensitive dependence of tomographic
peak analyses on $z_{{\rm bias}}$ than on $\sigma_{{\rm ph}}$ can
be understood as follow. The peak number counts depend on the redshift
through the lensing kernel, the redshift dependence of the mass function,
and the angular size of massive halos. If there is a $z_{{\rm bias}}$
in the photo-z measurements, it affects the peak counts in a systematic
way. Over (under) estimate the photo-z will lead to under (over) estimate
of $(\Omega_{{\rm m}},\sigma_{8})$. This leads to a very sensitive
dependence on $z_{{\rm bias}}$ for tomographic peak statistics. On
the other hand, for the dispersion $\sigma_{{\rm ph}}$, to the linear
order, we can write the peak count near $z_{0}$ as $N(z)\approx N(z_{0})+(\mathrm{d}N(z)/{\rm d}z)(z-z_{0})$.
Thus the change on the peak count from the plus and minus $z-z_{0}$
parts cancels out. In this case, we expect that the dependence of
peak counts on $\sigma_{{\rm ph}}$ is minimal. In reality, there
are higher order terms of $z-z_{0}$ in the expansion beyond the linear
term, thus the canceling is partial. The smaller the $\sigma_{{\rm ph}}$,
the more canceling occurs. In our 2-bin analyses, the constraint on
$\sigma_{{\rm ph}}$ purely from the peak counts is relatively large,
and we still see a certain dependence on the priors of $\sigma_{{\rm ph}}$
for the tomographic peak studies (top right panel of Fig.\ref{fig:phz_degradation-upper}).
In the 4-bin case, the peak count data themselves can give already
a tight constraint on $\sigma_{{\rm ph}}$, and thus the canceling
discussed above is more complete. Thus there is nearly no dependence
on the priors of $\sigma_{{\rm ph}}$ for the peak analyses (top right
panel of Fig.\ref{fig:phz_degradation-lower}). Similarly, the more
sensitive dependence on $z_{{\rm bias}}$ than on $\sigma_{{\rm ph}}$
also shows up in the tomographic two point correlation analyses \citep{2006ApJ...636...21M,2006MNRAS.366..101H}.}

The requirements on the photo-z calibrations using spec-z are also
discussed. For the 4-bin case with the degradation factor of $\sim1.5$,
we need $\sigma^{\text{prior}}\left(z_{\text{bias}}\right)\approx10^{-4}$,
which in turn requires $N_{\text{ spect }}\approx10^{4}$ in each
redshift calibration bin for the fiducial $\sigma_{\text{{\rm ph}}}=0.02$.
We note that the calibration requirements depend on the assumed photo-z
distributions. More realistic photo-z distributions taking into account
the redshift dependence of their error parameters beyond $(1+z)$
assumption may result better estimates about the desired accuracy
of photo-z error parameters, and thus more realistic requirements
about the spec-z measurements. Other redshift calibration methods
are also worth to be explored. Particularly, for some surveys, such
as Euclid\footnote{\url{https://www.euclid-ec.org/}} and \textbf{Chinese
Space Station Telescope (CSST) \citep{2011SSPMA..41.1441Z,2018cosp...42E3821Z,2018cosp...42E1039F}},
they will have both imaging surveys and slitless spectroscopic surveys.
Detailed studies on how their spec-z sample can help the photo-z calibration
for the WL galaxy sample, both in terms of direct calibrations and
using the correlation methods, are needed in order to fully explore
their cosmological potentials. For that, specific survey characteristics
and the error budget should be taken into account in the analyses.

\textbf{In this paper, we focus on photo-z errors. There are also
other sources of systematics that can affect weak lensing peak studies.
Among these, galaxy intrinsic alignments can have significant effects.
They not only can generate additional shape noise \citep{2007ApJ...669...10F},
but also can affect the peak signals from clusters of galaxies. The
latter is related to the level of cluster member contaminations to
the source galaxies and their alignments in the host clusters. In
addition, the uncertainties of dark matter halo properties, such as
the halo mass function and the halo density profiles and their triaxialities
\citep{2005ApJ...635...60T}, and the baryonic effects \citep{2019MNRAS.tmp.1821F,2019arXiv190511636W},
etc., can also induce systematic errors in cosmological studies from
weak lensing peak statistics. Importantly, these effects can degenerate
to some degree with the effects of photo-z errors. For comprehensive
understandings of these effects, much more detailed investigations
taking into account different systematics are needed, which will be
the major efforts in our future studies.}

\acknowledgements{}

\textbf{We appreciate the very constructive comments and suggestions
from the referee that help to improve our paper. }This research is
supported in part by the National Natural Science Foundation of China
under the grants 11333001 and 11653001 and XXXXXXXX. X. K. Liu acknowledges
the support from NSFC-11803028 and YNU Grant KC1710708. Q. Wang acknowledges
the Strategic Priority Research Program of Chinese Academy of Sciences
Grant No. XDC01000000 and the National Key Program for Science and
Technology Research and Development (2017YFB0203300).

\appendix{}

\section{The Gaussian approximation for likelihoods}

\begin{figure}[H]
\begin{centering}
\includegraphics[scale=0.3]{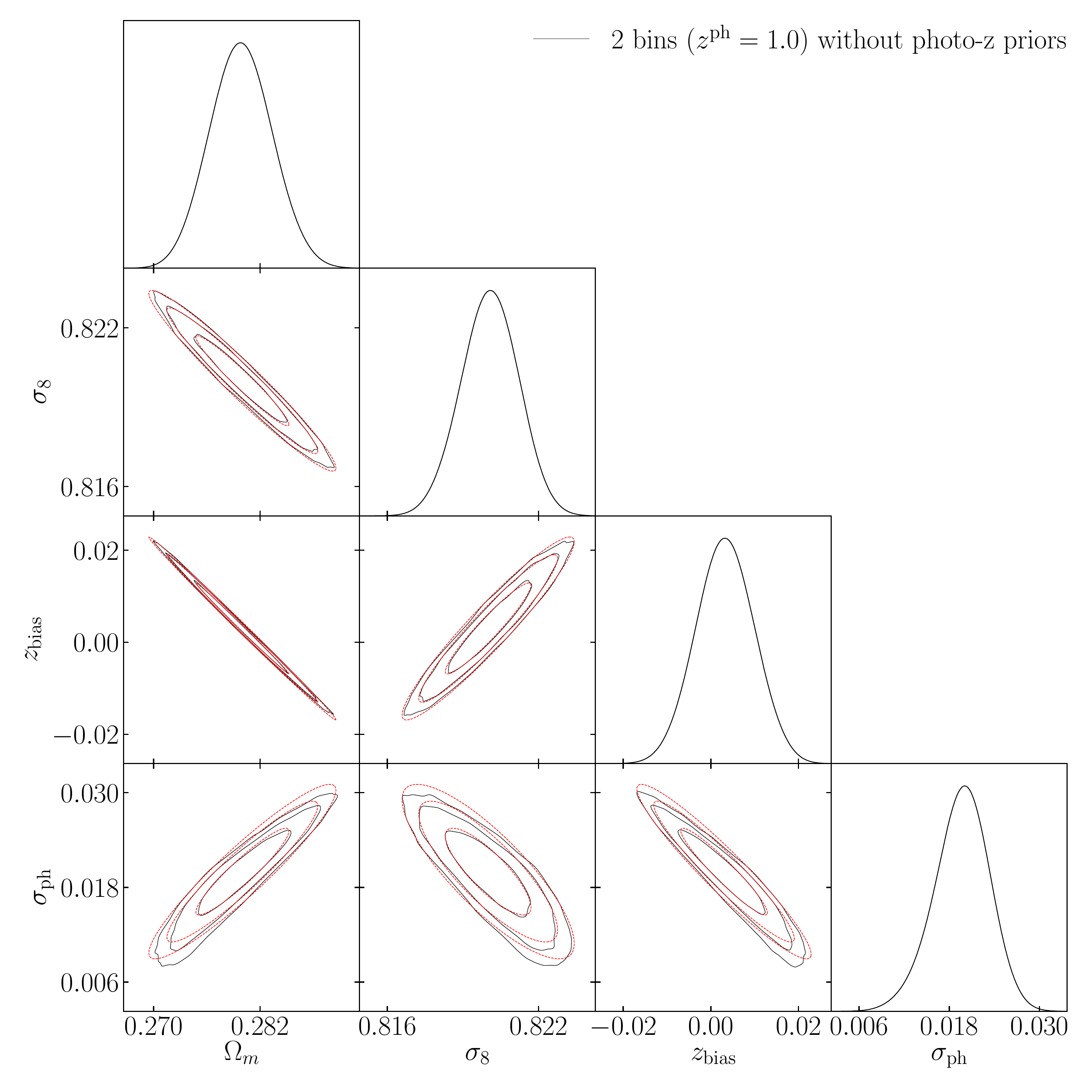}\includegraphics[scale=0.3]{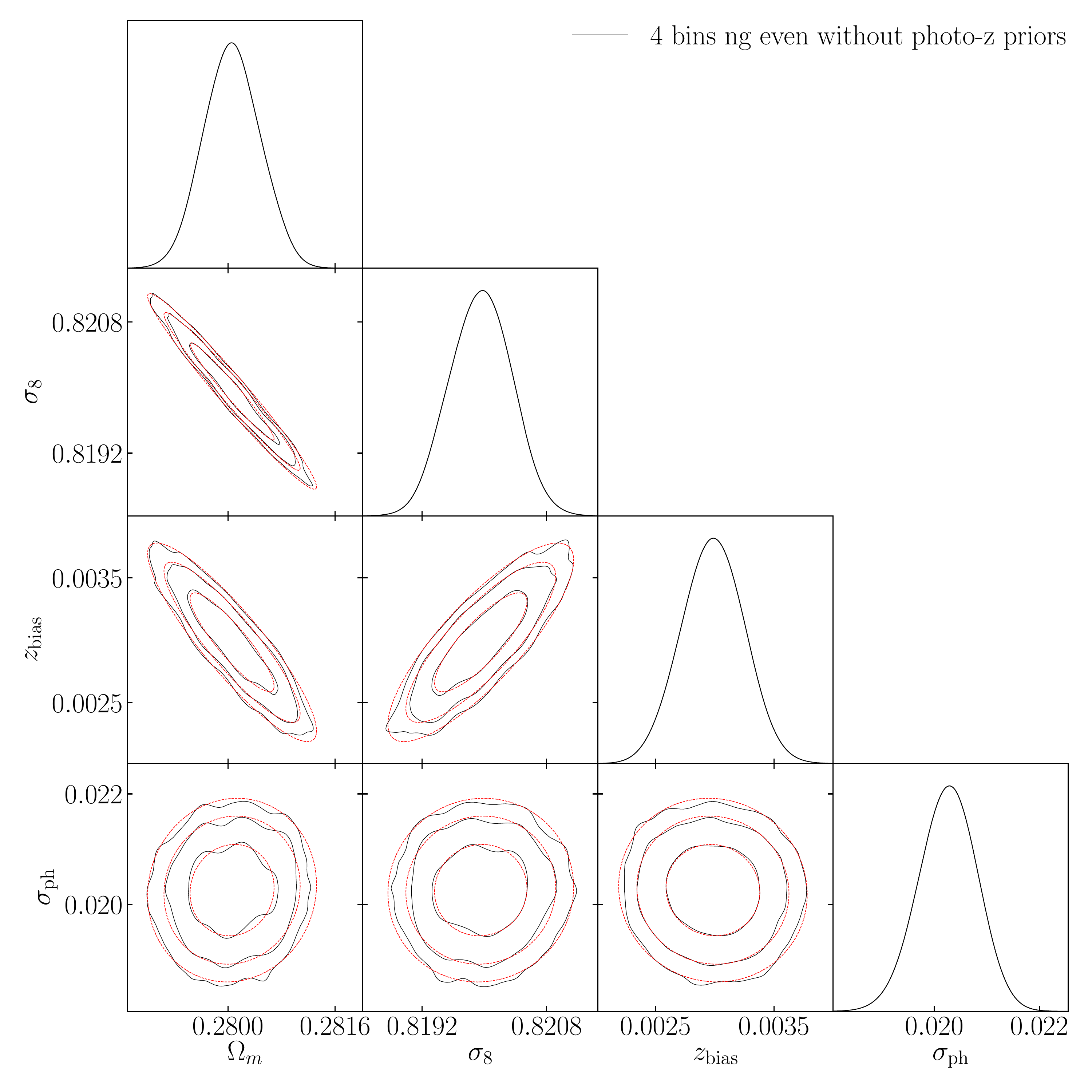}
\par\end{centering}
\figcaption{\label{fig:mcmc-fisher} Comparison of results from MCMC fitting (black)
and that from the Fisher approximation (red). Left and right are for
the 2-bin and 4-bin cases, respectively.}
\end{figure}

In Sec.5.4, we employ the Fisher approximation to study how the photo-z
errors affect the tomographic WL peak analyses and the derived cosmological
constraints. Here in Fig.\ref{fig:mcmc-fisher}, we show the comparisons
of the MCMC results (black) and those from Fisher approximation (red)
for the 2-bin and 4-bin cases considered in Sec.5.4. It is seen that
in both cases, the constraints from the Fisher approximation agree
excellently with those from MCMC fitting, showing its validity in
our forecast studies assuming LSST-like survey parameters.

\bibliographystyle{apj}
\bibliography{tom}

\clearpage
\end{document}